\documentclass[twocolumn]{aastex631}

\usepackage[english]{babel} 
\usepackage[T1]{fontenc}    
\usepackage{ae,aecompl}
\usepackage{pgf,pgfarrows,pgfnodes,pgfautomata,pgfheaps}
\usepackage{graphicx}       
\usepackage{natbib}         
\usepackage{url}            
\usepackage{grffile}        
\usepackage{mathtools}      
\usepackage{multirow}       
\usepackage{xspace}         

\usepackage{amsmath,amssymb,amsxtra,amsfonts}   

\newcommand{\lya}{\textrm{Ly}\ensuremath{\alpha}\xspace}

\usepackage{color}         
\definecolor{gold}{rgb}{1,0.80,0}
\definecolor{orange}{rgb}{1,0.5,0}
\definecolor{midgray}{gray}{0.3}
\definecolor{lblue}{rgb}{0,0.2,0.6}
\definecolor{dgreen}{rgb}{0.1,0.6,0.3}
\definecolor{purple}{rgb}{0.5019607843137255,0.0,0.5019607843137255}

\newcommand{\be}{\begin{equation}}
\newcommand{\ee}{\end{equation}}

\newcommand{\ba}{\begin{align}}
\newcommand{\ea}{\end{align}}

\newcommand{\defeq}{\vcentcolon=}

\newcommand{\Msun}{\ensuremath{M_\odot}\xspace}

\newcommand{\Mstar}{\ensuremath{M_\ast}\xspace}
\newcommand{\Lstar}{\ensuremath{L_\ast}\xspace}

\newcommand{\oh}{\ensuremath{12+\log({\rm O/H})}\xspace}

\newcommand{\fesc}{\ensuremath{f_{\rm esc}}\xspace}
\newcommand{\xion}{\ensuremath{\xi_{\rm ion}}\xspace}

\newcommand{\Muv}{\ensuremath{M_{\rm UV}}\xspace}

\newcommand{\K}{\ensuremath{\rm K}\xspace}

\newcommand{\Lunit}{\ensuremath{\rm erg~s^{-1}}\xspace}
\newcommand{\Funit}{\ensuremath{\rm erg~s^{-1}~cm^{-2}}\xspace}

\def\micron{\ensuremath{\mu\textrm{m}}\xspace}  

\newcommand\ionp[2]{#1$\;${\scshape{#2}}}      
\newcommand\ions[2]{#1$\;${\scshape{#2}}]}     
\newcommand{\Ha}{\textrm{H}\ensuremath{\alpha}\xspace}
\newcommand{\Hb}{\textrm{H}\ensuremath{\beta}\xspace}
\newcommand{\Hg}{\textrm{H}\ensuremath{\gamma}\xspace}
\newcommand{\Hd}{\textrm{H}\ensuremath{\delta}\xspace}
\newcommand{\HII}{\textrm{H}\textsc{ii}\xspace}

\newcommand{\OII}{[\textrm{O}~\textsc{ii}]\xspace}
\newcommand{\OIII}{[\textrm{O}~\textsc{iii}]\xspace}
\newcommand{\CIII}{\textrm{C}~\textsc{iii}]\xspace}

\newcommand{\HeII}{\textrm{He}~\textsc{ii}\xspace}
\newcommand{\CIV}{\textrm{C}~\textsc{iv}\xspace}

\newcommand{\NeIV}{[\textrm{Ne}~\textsc{iv}]\xspace}

\newcommand{\sersic}{S\'{e}rsic\xspace}
\def\U{\ensuremath{U_{360}}\xspace}     
\def\B{\ensuremath{B_{435}}\xspace}
\def\V{\ensuremath{V_{606}}\xspace}
\def\I{\ensuremath{I_{814}}\xspace}
\def\Y{\ensuremath{Y_{105}}\xspace}
\def\J{\ensuremath{J_{125}}\xspace}

\def\H{\ensuremath{H_{160}}\xspace}

\newcommand{\emc}{\textsc{Emcee}\xspace}

\newcommand{\galfit}{\textsc{Galfit}\xspace}

\newcommand{\bagp}{\textsc{BAGPIPES}\xspace}

\newcommand{\msa}{\textsc{msaexp}\xspace}
\newcommand{\ppxf}{\textsc{pPXF}\xspace}

\newcommand{\jwst}{\textit{JWST}\xspace}

\newcommand{\hst}{\textit{HST}\xspace}

\def\ie{i.e.\xspace}
\def\eg{e.g.\xspace}

\renewcommand\({\left(}
\renewcommand\){\right)}

\newcommand{\el}[1]{\ensuremath{\textrm{EL}_{#1}}}

\def\p{{\rm prior}}

\newcommand{\n}{\ensuremath{{\nu}\rm}\xspace}

\usepackage{cases}
\usepackage[flushleft]{threeparttable}

\newcommand{\galname}{RXJ2129-z8HeII\xspace}
\newcommand{\narrow}{0ex}
\newcommand{\Nline}{6\xspace}

\begin{document}

\title{
A strong He II $\lambda$1640 emitter with extremely blue UV spectral slope at $z=8.16$: presence of Pop III stars?
}

\author[0000-0002-9373-3865]{Xin Wang}
\affiliation{School of Astronomy and Space Science, University of Chinese Academy of Sciences (UCAS), Beijing 100049, China}
\affiliation{Institute for Frontiers in Astronomy and Astrophysics, Beijing Normal University, Beijing 102206, China}
\affiliation{National Astronomical Observatories, Chinese Academy of Sciences, Beijing 100101, China}

\author[0000-0003-0202-0534]{Cheng Cheng}
\affiliation{Chinese Academy of Sciences South America Center for Astronomy, National Astronomical Observatories, CAS, Beijing 100101, China}

\author[0000-0002-1971-5458]{Junqiang Ge}
\affiliation{National Astronomical Observatories, Chinese Academy of Sciences, Beijing 100101, China}

\author[0009-0006-0596-9445]{Xiao-Lei Meng}
\affiliation{National Astronomical Observatories, Chinese Academy of Sciences, Beijing 100101, China}

\author[0000-0002-3331-9590]{Emanuele Daddi}
\affiliation{Laboratoire AIM, CEA/DSM-CNRS-Universit\'e Paris Diderot, IRFU/Service d'Astrophysique, B\^at. 709, CEA Saclay, F-91191 Gif-sur-Yvette Cedex, France}

\author[0000-0001-7592-7714]{Haojing Yan}
\affiliation{Department of Physics and Astronomy, University of Missouri-Columbia, Columbia, MO 65211, USA}

\author[0000-0001-7673-2257]{Zhiyuan Ji}
\affiliation{Steward Observatory, University of Arizona, 933 North Cherry Avenue, Tucson, AZ 85721, USA}

\author[0000-0003-0401-3688]{Yifei Jin}
\affiliation{Institute for Theory and Computation, Harvard-Smithsonian Center for Astrophysics, Cambridge, MA 02138, USA}

\author[0000-0001-5860-3419]{Tucker Jones}
\affiliation{Department of Physics and Astronomy, University of California Davis, 1 Shields Avenue, Davis, CA 95616, USA}

\author[0000-0001-6919-1237]{Matthew A. Malkan}
\affiliation{Department of Physics and Astronomy, University of California Los Angeles, 430 Portola Plaza, Los Angeles, CA 90095, USA}

\author[0000-0002-7959-8783]{Pablo Arrabal Haro}
\affiliation{NSF's National Optical-Infrared Astronomy Research Laboratory, 950 N. Cherry Ave., Tucson, AZ 85719, USA}

\author[0000-0003-2680-005X]{Gabriel Brammer}
\affiliation{Cosmic Dawn Center (DAWN), Denmark}
\affiliation{Niels Bohr Institute, University of Copenhagen, Jagtvej 128, DK-2200 Copenhagen N, Denmark}

\author[0000-0003-3484-399X]{Masamune Oguri}
\affiliation{Center for Frontier Science, Chiba University, 1-33 Yayoi-cho, Inage-ku, Chiba 263-8522, Japan}
\affiliation{Department of Physics, Chiba University, 1-33 Yayoi-Cho, Inage-Ku, Chiba 263-8522, Japan}

\author[0000-0001-9062-8309]{Meicun Hou}
\affiliation{Kavli Institute for Astronomy and Astrophysics, Peking University, Beijing 100871, China}

\author{Shiwu Zhang}
\affiliation{Research Center for Astronomical Computing, Zhejiang Laboratory, Hangzhou 311100, China}

\correspondingauthor{Xin Wang}
\email{xwang@ucas.ac.cn}

\begin{abstract}
\noindent
Cosmic hydrogen reionization and cosmic production of first metals are major phase transitions of the universe occurring during the first billion years after the Big Bang, however these are still underexplored observationally. 
Using the \jwst NIRSpec prism spectroscopy, we report the discovery of a 
sub-\Lstar galaxy at $z_{\rm spec}=8.1623\pm0.0007$, dubbed \galname,
via the detection of a series of strong rest-frame UV/optical nebular emission lines and the clear Lyman break. \galname shows a pronounced UV continuum with an extremely steep (\ie blue) spectral slope of $\beta=-2.53_{-0.07}^{+0.06}$, the steepest amongst all spectroscopically confirmed galaxies at $z_{\rm spec}\gtrsim7$, in support of its very hard ionizing spectrum that could lead to a significant leakage of its ionizing flux.
Therefore, \galname is representative of the key galaxy population driving the cosmic reionization.
More importantly, we detect a strong \HeII$\lambda$1640 emission line in its spectrum, one of the highest redshifts at which such a line is robustly detected. 
Its high rest-frame equivalent width (${\rm EW}=21\pm4$ \AA) and extreme flux ratios with respect to UV metal and Balmer lines raise the possibility that part of \galname’s stellar populations could be Pop III-like.
Through careful photoionization modeling, we show that the physically calibrated phenomenological models of the ionizing spectra of Pop III stars with strong mass loss can successfully reproduce the emission line flux ratios observed in \galname.
Assuming the Eddington limit, the total mass of the Pop III stars within this system is estimated to be $7.8\pm1.4\times10^5\Msun$.
To date, this galaxy presents the most compelling case in the early universe where trace Pop III stars might coexist with metal-enriched populations.
\end{abstract}


\section{Introduction} \label{sect:intro}

The identification and characterization of the first generation of stars (Pop III) are of paramount importance as this quest provides key insights into the stellar evolution physics in the early Universe \citep{Bromm.2013}. The observational signatures expected from early galaxies dominated by these young, metal-free, massive Pop III stars are highly associated with very hard ionizing flux, given the high effective temperature of Pop III stars and a top-heavy initial mass function \citep{Nakajima.2022,Zackrisson.2023}.
This flux can ionize the interstellar medium (ISM) and produce strong hydrogen and helium emission lines, such as strong \lya, Balmer lines (\eg, \Ha, \Hb), $\HeII~\lambda1640~(\defeq\HeII)$ and \HeII$\lambda$4686, without the corresponding metal emission lines \citep{Schaerer.2002,Schaerer.2003,Zackrisson.2011}.
It is believed that Pop III galaxies with stellar mass $(\Mstar)\sim10^{4-5}\Msun$ consisting of a pure Pop III stellar population and no metal-enriched ones, are extremely hard to detect, even with \jwst, unless they are substantially magnified by gravitational lensing \citep{Windhorst.2018,Vikaeus.2022mn}. 
More massive Pop III galaxies with $\Mstar\gtrsim10^6\Msun$ can be first photometrically selected due to their blue UV continuum, and spectroscopically confirmed by their high ionization nebular emission lines and nebular continuum \citep{trusslerObservabilityIdentificationPopulation2023}.
Yet they should be really rare, with their characteristic nebular emission features fading within a short time scale of $\sim$10 Myrs at $z\sim8-9$ \citep{katzChallengesIdentifyingPopulation2023}.

Recent cosmological hydrodynamic simulations show that Pop III stars can continue to form also at later epochs, provided that pockets of primordial gas can be preserved during cosmic evolution. This is possible either in halos that gain their gas from regions not yet polluted by outflows from nearby star-forming galaxies, or in halos whose progenitors had a suppression of star formation \citep{Ciardi.2005,Venditti.2023}. This naturally leads to a type of ``hybrid'' Pop III galaxies, which have better chances to be identified using current facilities. One important clue of such ``hybrid'' Pop III stellar populations is the existence of the strong \HeII line, with rest-frame equivalent width (EW) as high as 100-150 \AA{}, powered by the hard ionizing background radiation from the Pop III stars.

At low redshifts, such broad nebular \HeII emission is very rare and its spectral properties are usually consistent with the ionizing sources of Wolf-Rayet stars, stripped stars, X-ray binaries, or active galactic nuclei \citep[see \eg,][]{Berg.2016,Senchyna.2019cos,Nanayakkara.2019,Saxena.2020}. In addition, this \HeII emission is usually accompanied by other prominent high-ionization lines such as \ionp{N}{v}$\lambda$1240, $\CIV\lambda\lambda1548,1551~(\defeq\CIV)$, and $\CIII\lambda\lambda1907,1909~(\defeq\CIII)$, of similar excitation energy \citep{Sobral.2015,starkSpectroscopicDetectionIv2015}.
Therefore, a clear path moving forward to identify Pop III stars in and beyond the epoch of reionization (EoR) is to search for pronounced \HeII emission without the accompanying rest-frame UV metal lines that are often observed in the low-redshift universe.

In this work, we present our detailed analysis of a galaxy (dubbed \galname) spectroscopically confirmed at $z_{\rm spec}=8.1623\pm0.0007$, showing strong \HeII emission and absence of any UV metal lines --- an excellent candidate likely having a mixture of Pop II and Pop III stars.
The overall physical properties of \galname are showcased in Table~\ref{tab:gal}.
This identification benefits from the synergy of the powerful \jwst spectroscopy and the lensing magnification boost from foreground massive cluster of galaxies.

This paper is orchestrated as follows. First in Sect.~\ref{sect:data}, we describe the novel JWST observations analyzed in this work. Then in Sect.~\ref{sect:lens} we briefly explain the lens model used to correct for the cluster magnification of our source. The details of our analysis methods and primary scientific results are given in Sect.~\ref{sect:rslt}. Finally, We summarize our main findings in Sect.~\ref{sect:conclu}. 
Throughout this paper, we adopt the standard concordance $\Lambda$CDM model with $\Omega_{\rm m} = 0.3$, $\Omega_{\rm \Lambda} = 0.7$, $H_{\rm 0} = 70$ km s$^{-1}$ Mpc$^{-1}$, and the AB magnitude system \citep{Oke.1983}.

{\small
\begin{table}
\centering
\begin{tabular}{lcccc}
 \hline\hline 
 Parameters  &  Values \\
 \hline \noalign {\smallskip}
R.A. [deg] & 322.416266 \\ [\narrow]
Decl. [deg] & 0.099675   \\ [\narrow]
$z_{\rm spec}$ &  8.1623$_{-0.0007}^{+0.0007}$ \\ [\narrow]
$\mu$ & $2.26 _{-0.14} ^{+0.14}$    \\
 \hline \noalign {\smallskip}
\multicolumn{2}{c}{Spectro-photometric analyses (Sect.~\ref{subsect:bagp})}   \\ [\narrow]
$\log(M_{\ast}/M_{\odot})$ &  7.75$_{-0.06}^{+0.06}$ \\ [\narrow]
$A^{\rm S}_{\rm V}$ [mag] &  0.12$_{-0.04}^{+0.04}$ \\ [\narrow]
$\log(Z_{\ast}/Z_{\odot})$ &  -0.94$_{-0.04}^{+0.07}$ \\ [\narrow]
$t_{\rm age}$ [Myr] &  216.38$_{-96.04}^{+68.89}$ \\ [\narrow]
$M_{{\rm UV}}$ [mag] &  -19.58$_{-0.02}^{+0.03}$ \\ [\narrow]
$\beta$ &  -2.53$_{-0.07}^{+0.06}$ \\ [\narrow]
$f^{\rm LyC}_{\rm esc}$ &  0.16$_{-0.03}^{+0.03}$ \\
 \hline \noalign {\smallskip}
\multicolumn{2}{c}{Emission line analyses (Sect.~\ref{subsect:EL})}   \\ [\narrow]
EW$_{\HeII}$ [\AA]      &  $21\pm4$          \\ [\narrow]
$f_{\HeII}/f_{\CIV}$    &  >2.1 (2-$\sigma$) \\ [\narrow]
$f_{\HeII}/f_{\CIII}$   &  >2.9 (2-$\sigma$) \\ [\narrow]
$f_{\HeII}/f_{\Hb}$     &  $1.7\pm0.4$       \\ [\narrow]
$f_{\OIII}/f_{\OII}$    &  $11.7\pm4.2$      \\ [\narrow]
$f_{\OIII}/f_{\Hb}$     &  $5.5\pm0.8$       \\ [\narrow]
12+log(O/H) &  7.63$_{-0.09}^{+0.14}$        \\ [\narrow]
SFR [$M_{\odot}$/yr] &  9.56$_{-1.70}^{+4.51}$ \\ [\narrow]
$A^{\rm N}_{\rm V}$ [mag] &  <0.6 (1-$\sigma$) \\ [\narrow]
$\log(\xi_{\rm ion} {[\rm erg^{-1} Hz]})$ &  25.55$_{-0.06}^{+0.05}$ \\
 \hline \noalign {\smallskip}
 \end{tabular}
\caption{Physical properties of \galname. The parameter values are given in median posterior constraints followed by 1-$\sigma$ confidence interval (CI). The lensing magnification is predicted from the up-to-date GLAFIC lens model (see Sect.~\ref{sect:lens}), which is accounted for in the estimates of these physical properties when necessary.
All reported line flux ratios and limits have been corrected for dust extinction.}
\label{tab:gal}
\end{table}
}

\section{Observations and Data Reduction} \label{sect:data}

\begin{figure*}
 \centering
 \includegraphics[width=0.9\textwidth,trim=4cm 0cm 4cm 0cm,clip]{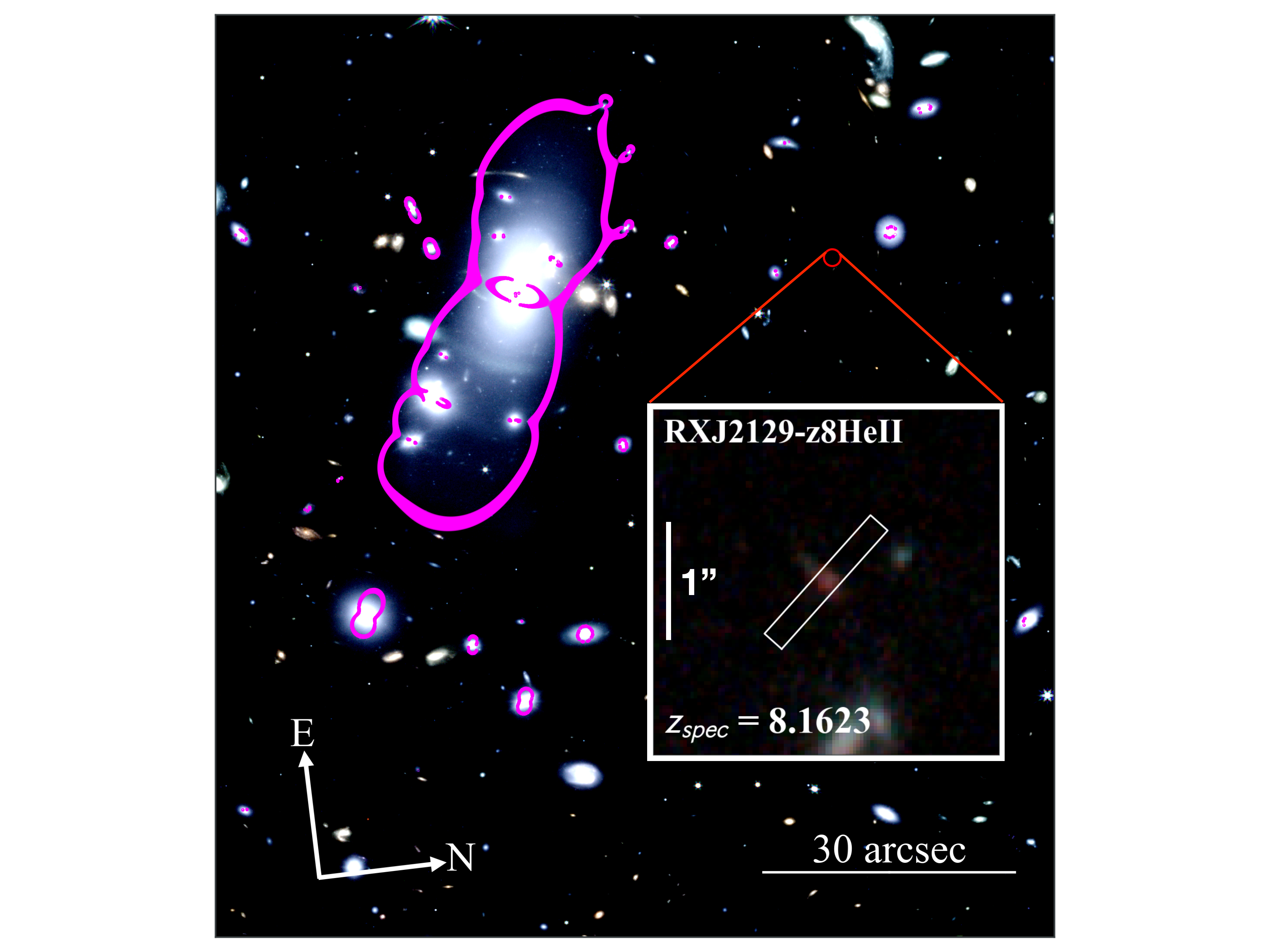}
 \vspace*{-1em}
 \caption{\small
 A color-composite NIRCam image of the RXJ2129 galaxy cluster. We use the F115W+F150W, F200W+F277W, and F356W+F444W imaging as the blue, green, and red colors, respectively. 
 The regions of formally infinite magnification at $z=8.16$ predicted by the up-to-date GLAFIC lens model \citep{Oguri.2010,Oguri.2021}, are represented by the magenta curves.
 The location of \galname is highlighted by the red circle. It is estimated to have a magnification factor of $\mu=2.26\pm0.14$ (see Sect.~\ref{sect:lens}).
 The inset shows a zoom-in view (3\arcsec$\times$3\arcsec) of \galname. Within this inset, we also show the orientation of the NIRSpec slit (0\farcs2~$\times$~1\farcs4), constructed by three MSA slitlets, through which the prism spectroscopy of \galname is taken by the \jwst Director’s Discretionary program (DD-2767;  PI: P. Kelly).
 \label{fig:rgb}}
\end{figure*}

The \jwst observations analyzed in this work were acquired by a Director’s Discretionary program (DD-2767; PI: P. Kelly), targeting the field of the galaxy cluster RXJ2129.7+0005 (henceforth RXJ2129) at $z=0.234$.
The primary goal of DD-2767 is to measure the light curves and spectra of a strongly lensed supernova SN 2022riv at $z=1.52$ \cite{Kelly.2022a,Kelly.2022b} discovered in the RXJ2129 field, by a \hst SNAP program (GO-16729; PI P. Kelly).
These observations consist of an imaging component using the Near-Infrared Camera (NIRCam) and a spectroscopic component using the Near-Infrared Spectrograph (NIRSpec), which are described in Sect.~\ref{subsect:nircam} and Sect.~\ref{subsect:nirspec}, respectively.

\subsection{\jwst/NIRCam data reduction and mosaicing} \label{subsect:nircam}

The imaging exposures were taken on 6 October 2022 (UT dates quoted throughout) using the F115W, F150W, F200W, F277W, F356W, and F444W filters, covering the wavelength range of $\lambda_{\rm obs}\in[1,5]\micron$.
The exposure times are 4982 seconds for the F150W/F356W filters, and are 2061 seconds for the other four NIRCam filters (F115W, F200W, F277W, and F444W), equivalent to a 5-$\sigma$ limiting depth of $\sim$29 mag.
We reduce these NIRCam images using the standard \jwst pipeline version 1.9.4 and making use of the {\sc jwst\_1040.pmap} context. We follow the standard three stages of the image data reduction pipeline \textsc{calwebb\_detector1}, \textsc{calwebb\_image2} and \textsc{calwebb\_image3} to calibrate each image and build the mosaic image of each band. Furthermore, we remove the ``snowball''\footnote{\url{https://jwst-docs.stsci.edu/data-artifacts-and-features/snowballs-artifact}} and the 1/f noise using the scripts from {\url{https://github.com/chriswillott/jwst}}. The ``wisps'' features\footnote{\url{https://jwst-docs.stsci.edu/jwst-near-infrared-camera/nircam-features-and-caveats/nircam-claws-and-wisps}} are removed using the default wisps templates\footnote{\url{https://stsci.app.box.com/s/1bymvf1lkrqbdn9rnkluzqk30e8o2bne}}. 
We produce the imaging mosaics at the 40 and 20 mas pixel scales, astrometrically aligned to the GAIA DR2 astrometry frame.

\begin{figure*}
 \centering
 \hspace{1cm}
 \includegraphics[width=\textwidth,trim=0cm 3.6cm 0cm 3cm,clip]{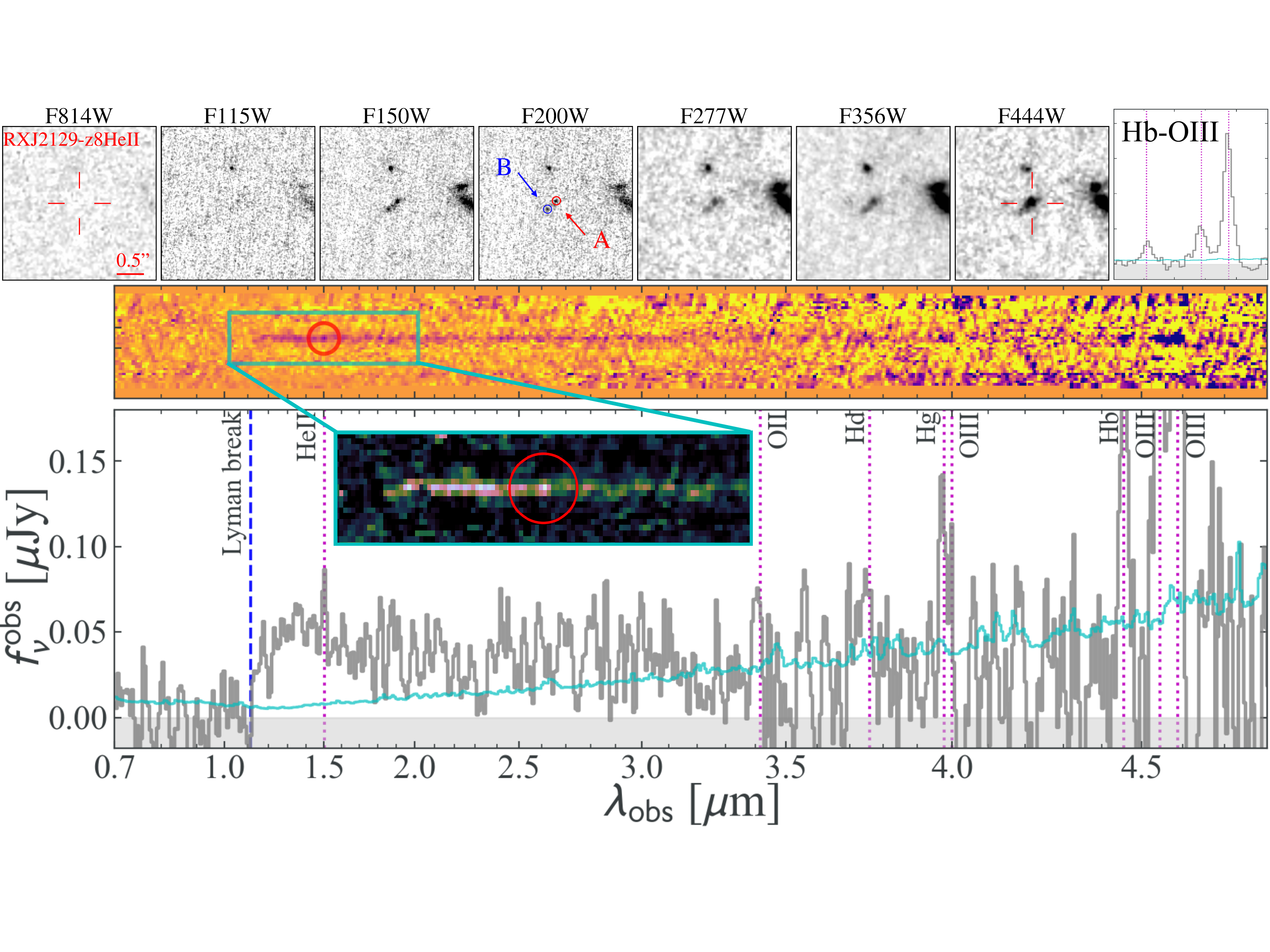}
 \vspace*{-1em}
 \caption{\small
 \jwst+\hst imaging and \jwst/NIRSpec spectroscopy of \galname.
 {\bf Top}: \galname's postage stamp images in multiple filters, cut from the \hst CLASH (PI: M. Postman) and \jwst/NIRCam (DD-2767;  PI: P. Kelly) imaging mosaics.
 These stamps are oriented such that north is up and east is to the left, which is different from the orientation in Figure~\ref{fig:rgb}.
 The strong flux deficit in ACS/F814W and NIRCam/F115W bands, as compared to the other NIRCam filters indicates $z\gtrsim8$.
 {The rightmost panel in the top row zooms in on the \OIII-\Hb line complex taken from the full NIRSpec prism 1D spectrum shown in the bottom row.}
 {\bf Middle and Bottom}: NIRSpec prism spectroscopy of \galname. In the middle trace, we show the \emph{unsmoothed} 2D prism spectra combined from the three individual slit-dithering sequence, using the up-to-date reduction software \msa (see Sect.~\ref{subsect:nirspec}).
 We detect prominent emission features of \OIII and \Hb at $\lambda_{\rm obs}\sim 4.5~\micron$, and strong continuum break blueward of \lya at $\lambda_{\rm obs}\sim 1.1~\micron$.
 On the bottom row, we show the 1D spectrum optimally extracted from the 2D spectral trace in gray histograms with the 1-$\sigma$ uncertainty in cyan.
 We mark the locations of multiple rest-frame UV and optical emission lines using the magenta dotted lines, and the Lyman break in the blue dashed line.
 The strong detections of \OIII+\Hb and the Lyman break pinpoints a secure spectroscopic redshift of \galname to be $z_{\rm spec}=8.1623\pm0.0007$.
 {\bf Bottom inset}: a zoom-in view on the 2D spectrum around the rest-frame UV wavelength range, where the location of the \HeII line is marked by the red circle.
 The 2D spectrum is shown in unit of signal-to-noise ratio (\ie $f_\nu^{\rm obs}/\sigma_\nu^{\rm obs}$) per spatial pixel, color-coded in the range of [-1, 10].
 \label{fig:obs}}
\end{figure*}

In Fig.~\ref{fig:rgb}, we show a color-composite image produced from these data, where our source of interest (\galname) is highlighted in the inset zoom-in panel. 
Our source in fact has two components A and B, with A dominating the total flux in all filters in the NIRCam long wavelength channel.
Hereafter unless otherwise specified, we use \galname to designate its component A, which is targeted by NIRSpec described below.
We also take advantage of the archival \hst imaging obtained by the Cluster Lensing and Supernova Survey\footnote{\url{https://archive.stsci.edu/prepds/clash/}} \citep[CLASH, PI: M. Postman,][]{2012ApJS..199...25P}.
CLASH provides the imaging data in 8 filters taken with ACS/WFC (F775W, F814W, and F850LP) and WFC3/IR (F105W, F110W, F125W, F140W, and F160W).
We compile a photometric catalog combining both the new NIRCam imaging and the existing \hst imaging in a self-consistent manner (see Sect.~\ref{subsect:photom}).
The image stamps of our target are displayed in the top row of Fig.~\ref{fig:obs}.

\subsection{\jwst/NIRSpec MSA spectroscopy and data reduction} \label{subsect:nirspec}

The follow-up \jwst spectroscopy was carried out on 22 October 2022, using the NIRSpec instrument in the Multi-Object Spectroscopy (MOS) mode. 
\galname was among the NIRSpec targets that were pre-selected based on their photometric redshifts \citep{Williams.2022}.
The \jwst NIRSpec observation was carried out using the prism disperser, which offered a very wide, continuous wavelength coverage of $\lambda_{\rm obs}\in[0.6,5.3]~\micron$, with the resolution of $R=\lambda/\Delta\lambda\sim50-400$.
In the inset of Fig.~\ref{fig:rgb}, we show a zoom-in view of \galname with the position of the MOS slit superposed.
The slit is composed of three shutters in the micro-shutter array (MSA), resulting in a total size of $\sim0\farcs2\times1\farcs4$.
The observation adopts the standard 3-point nodding pattern to facilitate background subtraction, and the total exposure time of 4464.2 seconds.

We reduce the NIRSpec MSA spectroscopic data following our customized procedures. Our reduction uses the context file {\sc jwst\_1040.pmap}, which contains up-to-date reference files for NIRSpec. This version of the reference files is considered to be a major update to the NIRSpec S flats, F flats, and readnoise files, which have been calibrated using in-flight data.
First of all, the level-1 \textsc{calwebb\_detector1} calibration pipeline is employed to reduce the raw exposures (\textsc{uncal.fits}) into count rate maps (CRMs, i.e., \textsc{rate.fits}), during which the detector artifacts and cosmic rays are flagged and removed. The resulting CRMs are visually inspected, and any remaining artifacts and bad pixels are manually masked.
We then use the custom reduction software \msa\footnote{\url{https://github.com/gbrammer/msaexp}} to perform the remaining steps of the reduction. It first preprocesses the rate images to equalize the pedestal of each exposure, identify and correct the ``snowball'' and 1/f noise features. Then it calls some specific modules from the standard level-2 \textsc{calwebb\_spec2} calibration pipeline to carry out the bulk of the data reduction. These modules include {\sc AssignWcs, Extract2dStep, FlatFieldStep, PathLossStep, PhotomStep}, which perform WCS initialization, 2D extraction of spectra, slit-level flat-fielding, path-loss correction, wavelength and flux calibration of each science exposure.
\msa subsequently subtracts background from each exposure using the two associated exposures at the other dithered positions from the nodding sequence, and drizzles the background-subtracted science exposures onto a common wavelength grid via inverse-variance weighting, with outlier rejection and bad pixel masking.
We adopt an oversampling rate of 2 for the output wavelength grid, to Nyquist-sample the line spread function in order to improve the sampling of the emission line profiles.
Finally \msa obtains the 1D spectrum from the 2D combined spectral traces following an optimal extraction methodology \citep{Horne.1986}, which uses the actual light profile of our target as the optimal aperture for spectral extraction.
The object light profile along the cross-dispersion direction is modeled as a Gaussian function fit to the collapsed 2D spectra (summed along the dispersion direction).

As a result, we obtain the 2D/1D prism spectra, as shown in the middle/bottom panels of Fig.~\ref{fig:obs}, covering an uninterrupted, wide wavelength range of $\lambda_{\rm obs}\in[0.6, 5.3]~\micron$. For $z\sim8$ galaxies, this corresponds to a contiguous coverage of the rest-frame UV and optical spectral energy distribution (SED) at $\lambda_{\rm rest}\in[700, 6000]$ \AA.
We detect pronounced emission features of the \OIII doublets and \Hb line complex at $\lambda_{\rm obs}\sim4.5~\micron$, and a clear continuum break at $\lambda_{\rm obs}\sim1.1~\micron$ with prominent continuum flux redward of the break. The identification of this continuum break as the Lyman break occurring at the \lya wavelength ($\lambda_{\rm rest}=1216$ \AA) is in excellent agreement with the unambiguous \OIII+\Hb emission feature in the prism spectrum, leading to an accurate spectroscopic confirmation of \galname at $z_{\rm spec}=8.1623\pm0.0007$, when the Universe is $\sim$613 Myr years old.
This puts \galname deep in the EoR, when the IGM was mostly neutral hydrogen \cite{Fan.2006araa,Stark.2016}.

We also perform extensive tests to verify the calibration of NIRSpec prism spectra. On one hand, we cross-check the flux levels of our extracted spectra and that of our broad-band photometry from NIRCam imaging in the similar wavelength regime. They show good agreement with each other (within 10\%; see Fig.~\ref{fig:fit}).
On the other hand, we conduct detailed investigation of the wavelength calibration by performing line identifications in low-$z$ galaxies showing multiple emission lines across their entire prism spectra, observed in the same MSA mask and extracted in the same fashion.
We find that the wavelength offset between the best-fit centroid and that expected for the mean redshift shows a scatter <0.004~$\micron$ in the observer frame, much smaller than the instrument resolution. So we conclude that our flux and wavelength calibrations are sufficiently accurate.

\section{Strong lensing models} \label{sect:lens}

We adopt the cluster lens model of RXJ2129 constructed using the GLAFIC software \citep{Oguri.2010,Oguri.2021}, updated from the initial work of \citet{Okabe.2020}. In total, 22 individual images of 7 multiply lensed background galaxies spectroscopically confirmed by \citet{Caminha.2019} are used as the strong lensing constraints. The macroscopic mass model consists of one cluster-scale dark matter halo in the elliptical Navarro-Frenk-White profile \citep{Navarro.1997}, combined with galaxy-scale halos modeled using the pseudo-Jaffe profile according to the scaling relation of the velocity dispersion and luminosity of cluster member galaxies \citep{Kawamata.2017}. The best-fit model is derived using the Markov chain Monte Carlo sampling process with a $\chi^2$ minimization assuming a positional uncertainty of 0.4$"$. A hundred additional realizations are also created to bootstrap the 1-$\sigma$ statistical uncertainties for magnifications. As a consequence, we obtain the best-fit and 1-$\sigma$ CI of the magnification estimates of \galname to be 2.26 and [2.12, 2.40]. We also double check with an independent lens model built by the Lenstool software \citep{Caminha.2019}, and derive consistent results.

\section{Analysis and Results} \label{sect:rslt}

\subsection{\jwst/NIRCam and \hst/ACS-WFC3 photometry} \label{subsect:photom}

\begin{figure*}
 \centering
 \includegraphics[width=.8\textwidth,trim=0cm 0cm 0cm 0cm,clip]{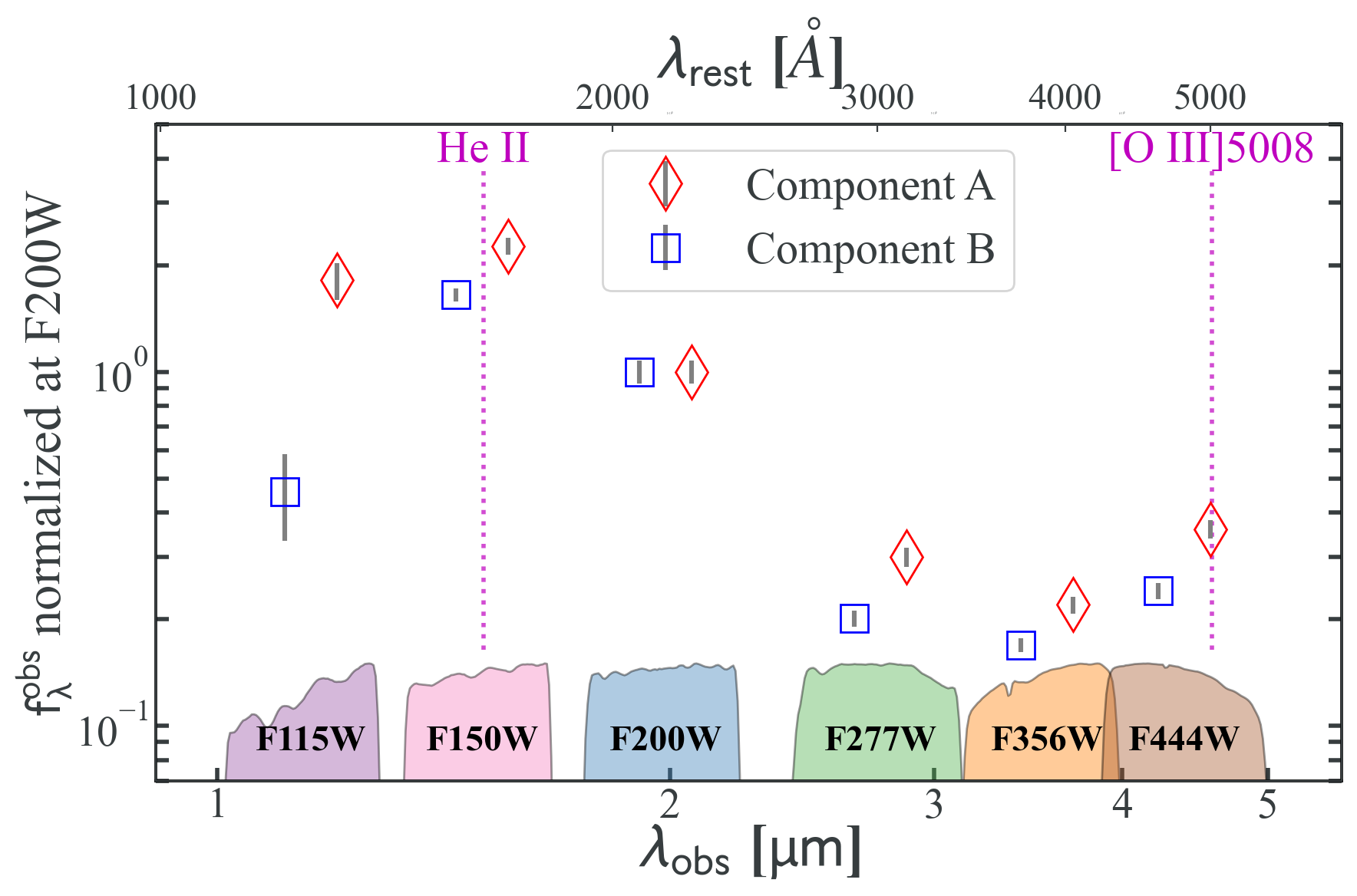}
 \vspace*{-1em}
 \caption{\small 
  We compare the photometry for both components of \galname, normalized at their individual F200W flux. In each filter, the data points are slightly offset in wavelength for clearer visualization.
  The photometry suggests that component B has comparably blue UV continuum, yet possibly with substantially fainter oxygen lines in the rest-frame optical (probed by F444W), indicating that component B is more metal poor than A. Also, component B's SED shape might accommodate \HeII with similar equivalent width, albeit at lower confidence.
  }
 \label{fig:ABphot}
\end{figure*}

For comprehensive photometry of \galname, we not only use the 6-band imaging mosaics produced from JWST NIRCam data, but also include the publicly released 8-band \hst imaging mosaics from CLASH \citep{2012ApJS..199...25P}. We first transform the CLASH 0.065$"$ imaging mosaics to mosaics on 0.04$"$ plate scale, and then PSF-match the 8 \hst ACS/WFC3 filters and 6 \jwst/NIRCam filters to the F444W resolution (FWHM = 0.14$"$). We utilize a window function to remove the high-frequency noise in the Fourier domain. We modify the rms images in order to match the F444W resolution for every other filters using $\rm RMS_{F444W, matched} = \sqrt{\rm \Sigma W_{i, F444W}^{2} \ast RMS_{i}^{2}} $, where $\rm RMS_{F444W, matched}$ and $\rm RMS_{i}$ are the rms after PSF-matching and the original rms in filter i respectively,
$\rm W_{i, F444W}$ is the kernel used to match all PSFs to that of the F444W filter through $\rm PSF_{F444W, matched} = PSF_{i} \ast W_{i, F444W}$.

{\small
\begin{table}
\centering
\begin{tabular}{lcccc}
 \hline\hline 
 Imaging Filters  &  \multicolumn{2}{c}{Observed Magnitudes [ABmag]} \\
  & Component A  & Component B \\
 \hline \noalign {\smallskip}
ACS/F775W   &  <26.32   &  <27.72 \\ [\narrow]
ACS/F814W   &  <27.57   &  <27.91 \\ [\narrow]
ACS/F850LP  &  <27.13   &  <28.00 \\ [\narrow]
WFC3/F105W  &  <27.57   &  <28.49 \\ [\narrow]
WFC3/F110W  &  27.41 $\pm$ 0.26     &  28.21 $\pm$ 0.52 \\ [\narrow]
WFC3/F125W  &  26.84 $\pm$ 0.28     &  <27.75 \\ [\narrow]
WFC3/F140W  &  26.84 $\pm$ 0.27     &  <27.83 \\ [\narrow]
WFC3/F160W  &  26.62 $\pm$ 0.20     &  <27.84 \\ [\narrow]
NIRCam/F115W  &  27.22 $\pm$ 0.31   &  <28.64 \\ [\narrow]
NIRCam/F150W  &  26.51 $\pm$ 0.12   &  27.57 $\pm$ 0.20 \\ [\narrow]
NIRCam/F200W  &  27.12 $\pm$ 0.20   &  27.50 $\pm$ 0.33 \\ [\narrow]
NIRCam/F277W  &  26.85 $\pm$ 0.13   &  28.53 $\pm$ 0.23 \\ [\narrow]
NIRCam/F356W  &  26.80 $\pm$ 0.11   &  28.16 $\pm$ 0.20 \\ [\narrow]
NIRCam/F444W  &  26.08 $\pm$ 0.11   &  27.32 $\pm$ 0.23 \\
 \hline \noalign {\smallskip}
 \end{tabular}
\caption{Detailed photometry of \galname separating components A and B from \hst ACS optical, WFC3 infrared, and \jwst NIRCam infrared imaging. All uncertainties presented here refer to the 1-$\sigma$ standard deviations. Upper limits are at 2-$\sigma$ confidence levels.
}
\label{tab:BBphot}
\end{table}
}

In the \jwst NIRCam F200W image, \galname shows two components, ``A'' and ``B'', which are marked by the arrows in the F200W stamp shown in Fig.~\ref{fig:obs}. The \jwst NIRSpec MOS slit falls on component A, which is also predominant in the total flux in the long-wavelength NIRCam filters. We use the F200W image as the detection image and obtain photometry for the two components separately. To maximize the detection SNRs, fluxes in each filter are measured within fixed apertures of 2$\times$FWHM of the F444W image ($0.28''$) in diameter. Furthermore, the flux is dust corrected for galactic extinction through $F_{\rm i,cor.} = F_{\rm i} \times 10^{(A_{\rm i}/2.5)}$, where $A_{\rm i}$ is the extinction in the $\rm i$-th band \citep{Cardelli.1989}.
The resultant photometric measurements of the two components of \galname is presented in Table~\ref{tab:BBphot}. For filters with no detections, we report the 2-$\sigma$ upper limits. The NIRCam photometry normalized at their individual F200W flux is shown in Fig.~\ref{fig:ABphot}.

\subsection{Photometric redshift estimates} \label{subsect:photoz}

\begin{figure*}
 \centering
 \includegraphics[width=.47\textwidth,trim=0cm 0cm 0cm 0cm,clip]{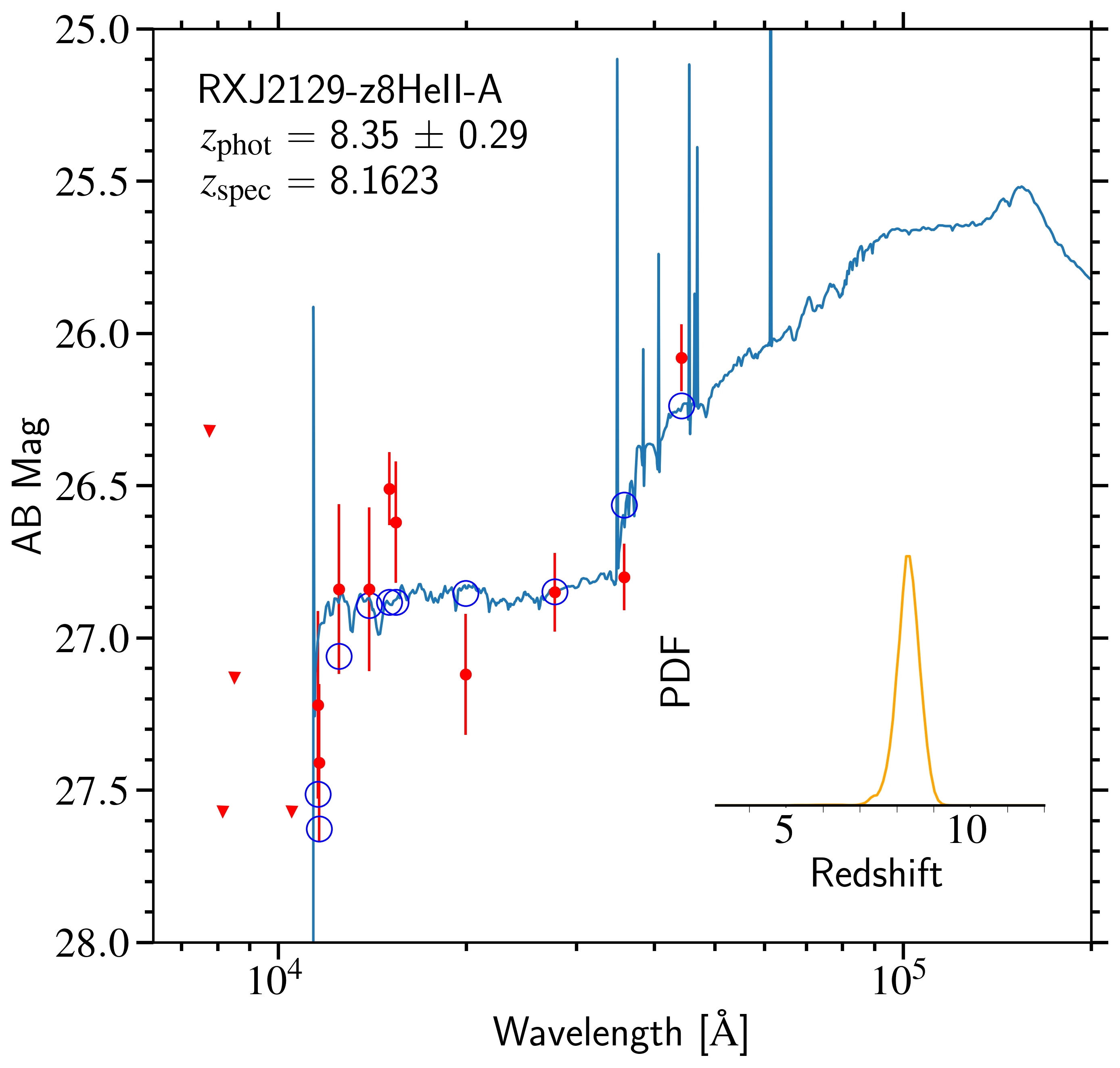}
 \includegraphics[width=.45\textwidth,trim=0cm 0cm 0cm 0cm,clip]{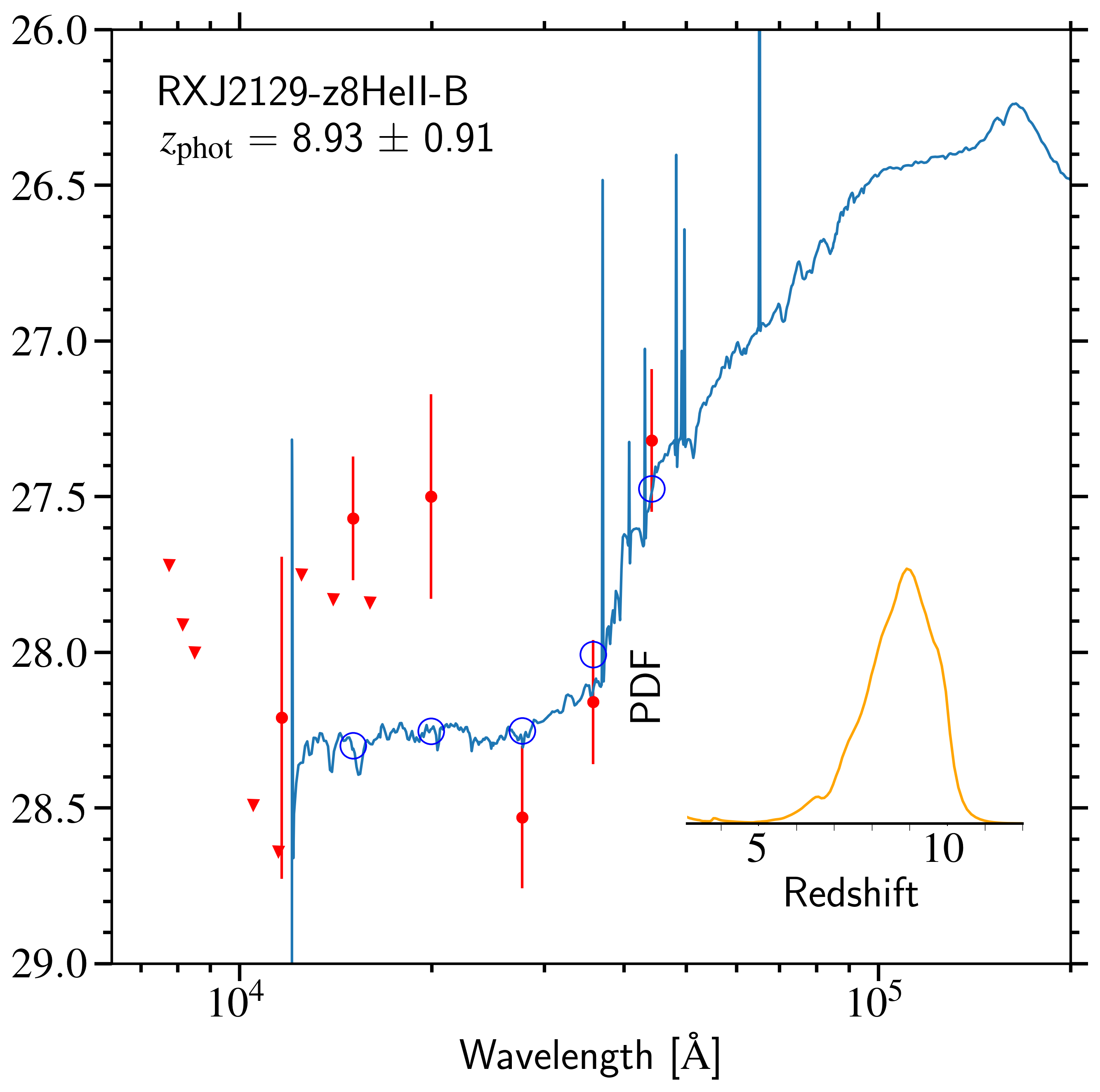} \\
 \vspace*{-1em}
 \caption{\small 
  \textsc{EAZY} fit to the broad-band photometry of \galname, performed on the components A and B separately. 
  The red points with error bars correspond to observed magnitudes with 1-$\sigma$ measurement uncertainties, whereas the triangles represents 2-$\sigma$ upper limits, 
  all given in Table~\ref{tab:BBphot}.
  The blue curve in each panel represents the best-fit SED model, with the open circles showing the model photometry in respective filters.
  The inset panel shows the probability distribution function of the photometric redshift, \ie, $P(z)$. The photometric redshift constraints of the two components are both in good agreement with $z_{\rm spec}=8.1623$ determined from the NIRSpec prism spectroscopy of component A. This strongly suggests that component B is also part of the galaxy \galname.
  }
 \label{fig:photoz}
\end{figure*}

We use the \texttt{EAzY} software \citep{Brammer.2008} to estimate the photometric redshift from the broad-band photometry presented in Table~\ref{tab:BBphot}. All together, we use the \hst filters ACS/F775W, ACS/F814W, ACS/F850LP, WFC3/F105W, WFC3/F110W, WFC3/F125W, WFC3/F140W, WFC3/F160W, as well as the \jwst NIRCam filters F115W, F150W, F200W, F277W, F356W and F444W. We adopt the standard set of galaxy SED templates {\sc eazy\_v1.1\_lines.spectra.param}, which includes star-forming galaxies with strong emission lines. The resulting photometric redshifts of components A and B are $z_{\rm phot}=8.35\pm0.29$ and $z_{\rm phot}=8.93\pm0.91$, respectively (see Fig.~\ref{fig:photoz}), in good agreement with the spectroscopic redshift that we derive from the NIRSpec prism spectroscopy.
{ As shown in Fig.~\ref{fig:photoz}, the model and observed photometry for the component A of \galname shows high consistency, except in filters NIRCam/F150W and WFC3/F160W, indicating strong emission features at $\lambda_{\rm obs}\sim1.5\micron$. The component B shows similar excess in observed photometry in the bluest NIRCam bands.
}

\subsection{Source morphology} \label{subsect:morph}

\begin{figure*}
 \centering
 \includegraphics[width=.8\textwidth,trim=0cm 1.5cm 0cm 1.5cm,clip]{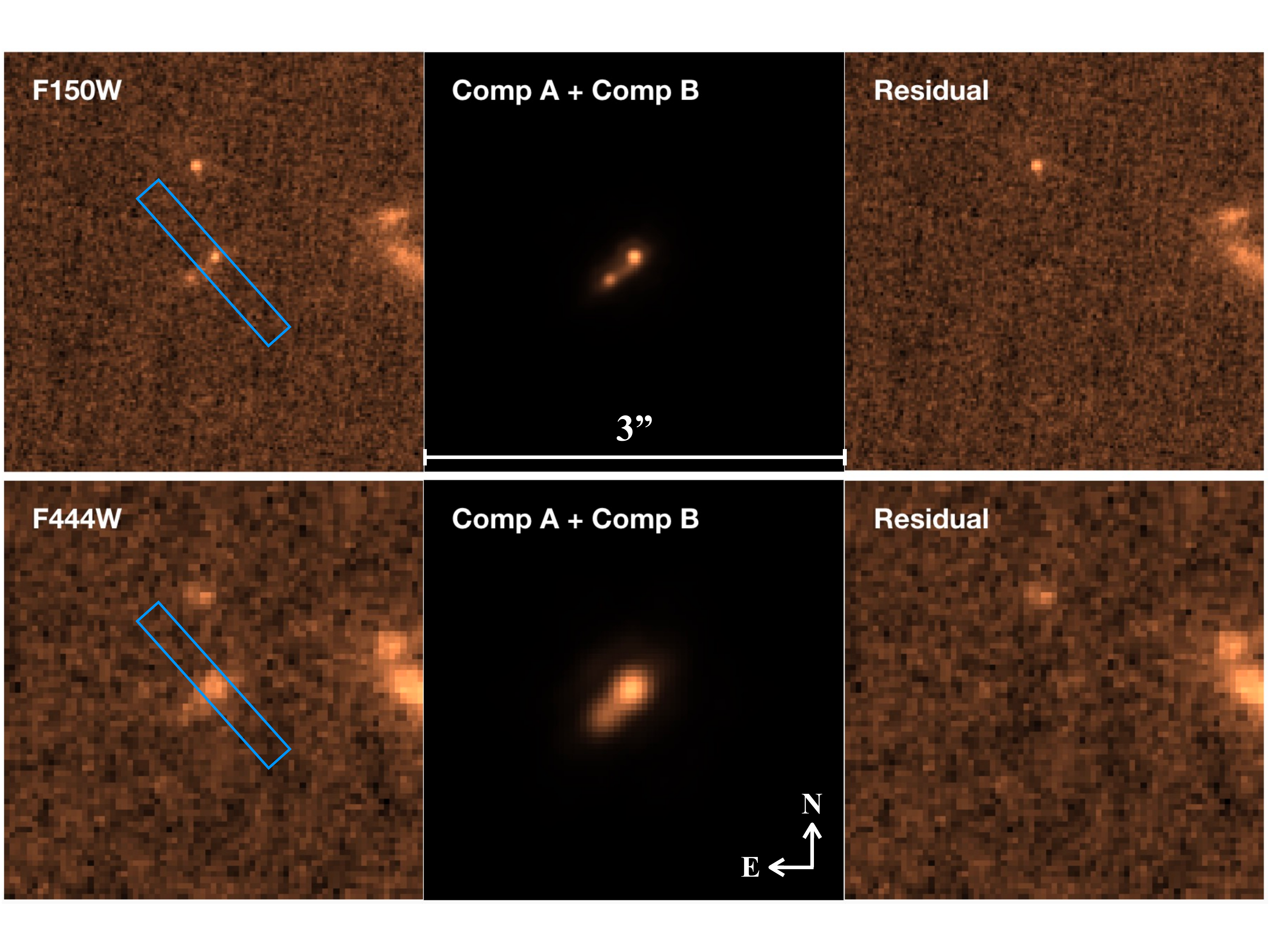}
 \caption{\small
    The rest-frame UV and optical morphology of the entire \galname galaxy, comprising two components: A and B. We use \galfit to perform morphological modeling of both components in \jwst NIRcam imaging taken at filters F150W (upper panels) and F444W (lower panels). At $z_{\rm spec}=8.16$, F150W and F444W correspond to rest-frame UV and optical wavelength ranges, respectively. In each row, we show the original observed image, \galfit models of both components, residual removing their \galfit models.
    On the observed images, we overlay the NIRSpec MSA slit represented by the blue box. The size of all stamps is 3"$\times$3", with north up and east to the left.
    As indicated by the residual panels, we achieve reasonable morphological models of the entire galaxy.
 \label{fig:morph}}
\end{figure*}

As shown in the multiple image stamps displayed in the upper panels of Fig.~\ref{fig:obs}, the entire \galname galaxy consists of two components: A and B, with the former dominating the flux in rest-frame optical probed by the NIRCam long wavelength channels, and the latter clearly manifesting in the rest-frame UV covered by the short wavelength channels, especially F150W.
Therefore, our morphology analyses are performed in two NIRCam bands, namely, F150W and F444W, representing the rest-frame UV and optical light profiles of our galaxy, respectively.
We use \galfit \citep{2002AJ....124..266P} to model both components in F150W. 
First, two models are chosen to fit component A simultaneously.
One is a \sersic profile describing the nucleated emission in the center with a \sersic index ($n$) of 4, an effective radius ($R_{\rm e}$) of $0\farcs048$, an axis ratio ($b/a$) of 0.85 and a position angle ($\theta$) of -48.8 deg. An exponential disk model is adopted to fit the underlying extended structure with disk scale-length ($R_{\rm s}$) of $0\farcs08$, $b/a$ of 0.26 and $\theta$ of -50.4 deg.
Component B is well reconstructed with a \sersic profile ($n=4$, $R_{\rm e}=0\farcs047$, $b/a=0.86$, $\theta$ = -50.8 deg) and an exponential disk model ($R_{\rm s}=0\farcs065$, $b/a=0.28$, $\theta$ = -48.7 deg).
The resulting model residual removing both components is shown in the upper right panel of Fig.~\ref{fig:morph}.

The inset of Fig.~\ref{fig:rgb} shows the position of the NIRSpec MOS slit with respect to \galname, which primarily covers its component A.
As implied by the observed prism spectra shown in the upper panel of Fig.~\ref{fig:obs}, the broad-band flux of the NIRCam F444W filter is likely dominated by the high equivalent width \OIII+\Hb nebular emission lines (also see Table~\ref{tab:ELflux}). Indeed we observe a more extended light profile in the rest-frame optical than that in the UV.
We thus model the image of component A in F444W with an exponential disk and a \sersic profile. 
The diffuse and extended structure is well reproduced by an exponential disk model with $R_{\rm s}$ of $0\farcs12$, $b/a$ of 0.3 and $\theta$ of -45.8 deg. The nucleated structure is modeled by a \sersic profile ($n=4$) with $R_{\rm e} = 0\farcs068$, $b/a=0.88$ and $\theta$ = 5.9 deg.
Component B in F444W is very faint and can be well fitted by an exponential disk model with $R_{\rm s}$ of $0\farcs04$, $b/a$ of 0.35 and $\theta$ of -40.9 deg.
The resulting model residual is shown in the lower right panel of Fig.~\ref{fig:morph}.

\subsection{Spectro-photometric analyses} \label{subsect:bagp}

\begin{figure*}
 \centering
 \includegraphics[width=.9\textwidth,trim=0cm 0cm 0cm 0cm,clip]{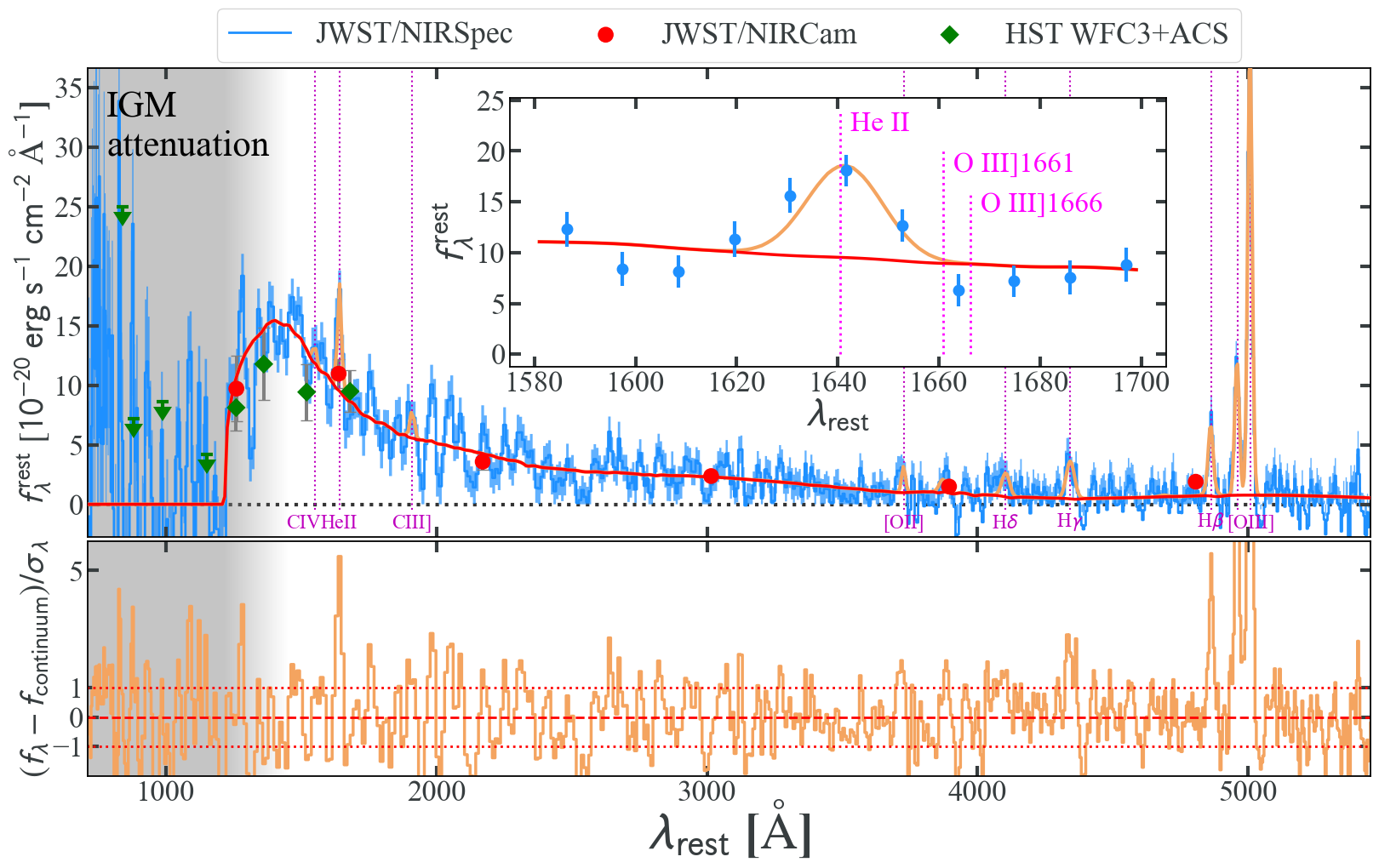}
 \caption{\small
    Spectro-photometric and emission line analyses of the full \jwst-\hst observations of \galname. 
    {\bf Top}: the optimally extracted 1D NIRSpec prism spectrum is shown in blue histograms with 1-$\sigma$ error spectrum in cyan bands. The broad-band photometric results from NIRCam and WFC3+ACS are represented by the red circles and green diamonds, respectively.
    We perform full spectrum analyses and fit Gaussian profiles to nebular emission features to estimate line fluxes and upper limits.
    The resulting best-fit continuum model and emission lines are represented by the red and orange curves. The top inset panel zooms in on the \HeII line, with significant detection.
    {\bf Bottom}: the ratio between the continuum-subtracted $f_{\lambda}$ flux and the measured 1-$\sigma$ error spectrum. For the wavelength ranges without significant emission line flux contribution, this ratio follows a $\mathcal{N}(0,1)$ distribution, implying that the noise properties are well defined. As clearly shown here, the most pronounced emission features in the rest-frame UV and optical wavelengths are \HeII and \OIII-\Hb lines, respectively.
    The gray shaded region marks the wavelength range affected by strong attenuation by the highly neutral IGM at $z\sim8$.
    \label{fig:fit}}
\end{figure*}

We employ multiple independent methods to conduct detailed spectro-photometric analyses of both the broad-band photometry and the full NIRSpec spectrum that we obtained for \galname.
Performing full spectrum fitting is highly critical since it is the only appropriate approach to extract the detailed information of stellar population from the low-resolution prism spectroscopy.

We first employ the \bagp software \citep{Carnall.2018} to perform SED fitting of both our spectroscopic and photometric data simultaneously. We adopt a widely used, double power law star-formation history model to describe the evolution of the cosmic star-formation rate (SFR) density \citep{Madau.2014}, which has the flexibility to account for both rising and declining star-formation activities:
\begin{align}
    {\rm SFR}(t) \propto \left[ \left(\frac{t}{\tau}\right)^{a} + \left(\frac{t}{\tau}\right)^{-b} \right]^{-1}.
\end{align}
It has a Jeffery's prior on the exponents: $a,b\in(0.01,1000)$ and a flat prior on the peak time of star formation: $\tau\in(0,t_{\rm H})$, where $t_{\rm H}$ is the Hubble time at the observed redshift. \bagp relies on the BC03 \citep{Bruzual.2003} stellar population synthesis model and the nebular emission model created by the \textsc{Cloudy} photoionization code \citep{ferland2017ReleaseCloudy2017}.
For other key assumptions, we choose the Chabrier initial mass function (IMF) \citep{Chabrier.2003}, the Calzetti dust attenuation law \citep{Calzetti.2000} with $A^{\rm S}_{\rm V}\in(0, 3)$, a stellar metallicity range of $Z/Z_{\odot}\in(0, 2)$, and a conservative Gaussian redshift prior of $z\sim N(8.16,0.01)$.

\begin{figure*}
 \centering
 \includegraphics[width=\textwidth,trim=0cm 0cm 0cm 0cm,clip]{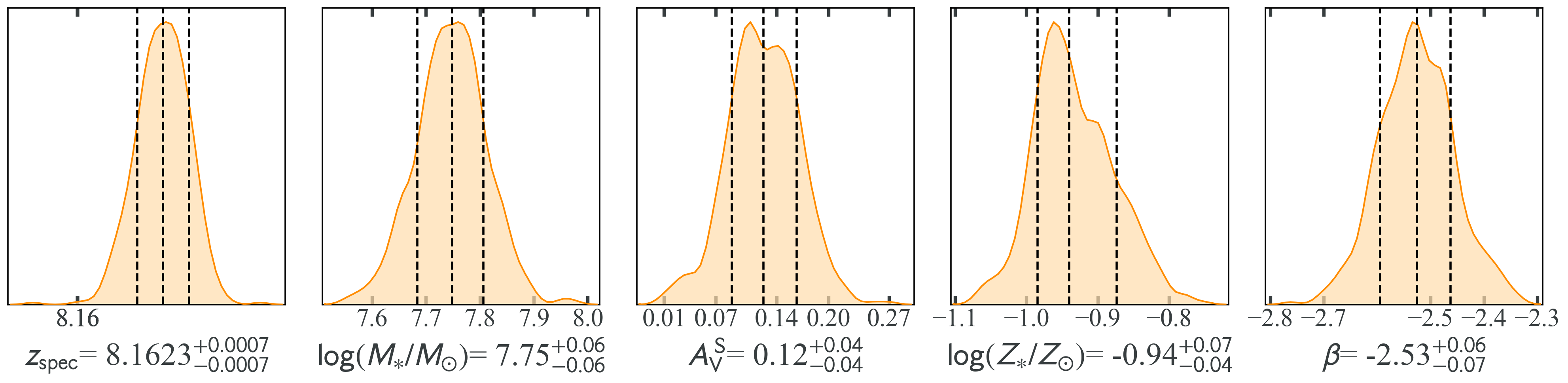} \\
 \caption{\small
 The 1D posterior distributions of some key parameters fit from our spectro-photometric analyses: spectroscopic redshift ($z_{\rm spec}$), stellar mass (\Mstar, already corrected for lensing magnification), stellar dust attenuation ($A^{\rm S}_{\rm V}$), stellar-phase metallicity ($Z_{\ast}$), and UV spectral slope ($\beta$).
 }
 \label{fig:post}
\end{figure*}

\begin{figure*}
 \centering
 \includegraphics[width=.8\textwidth,trim=0cm 0cm 0cm 0cm,clip]{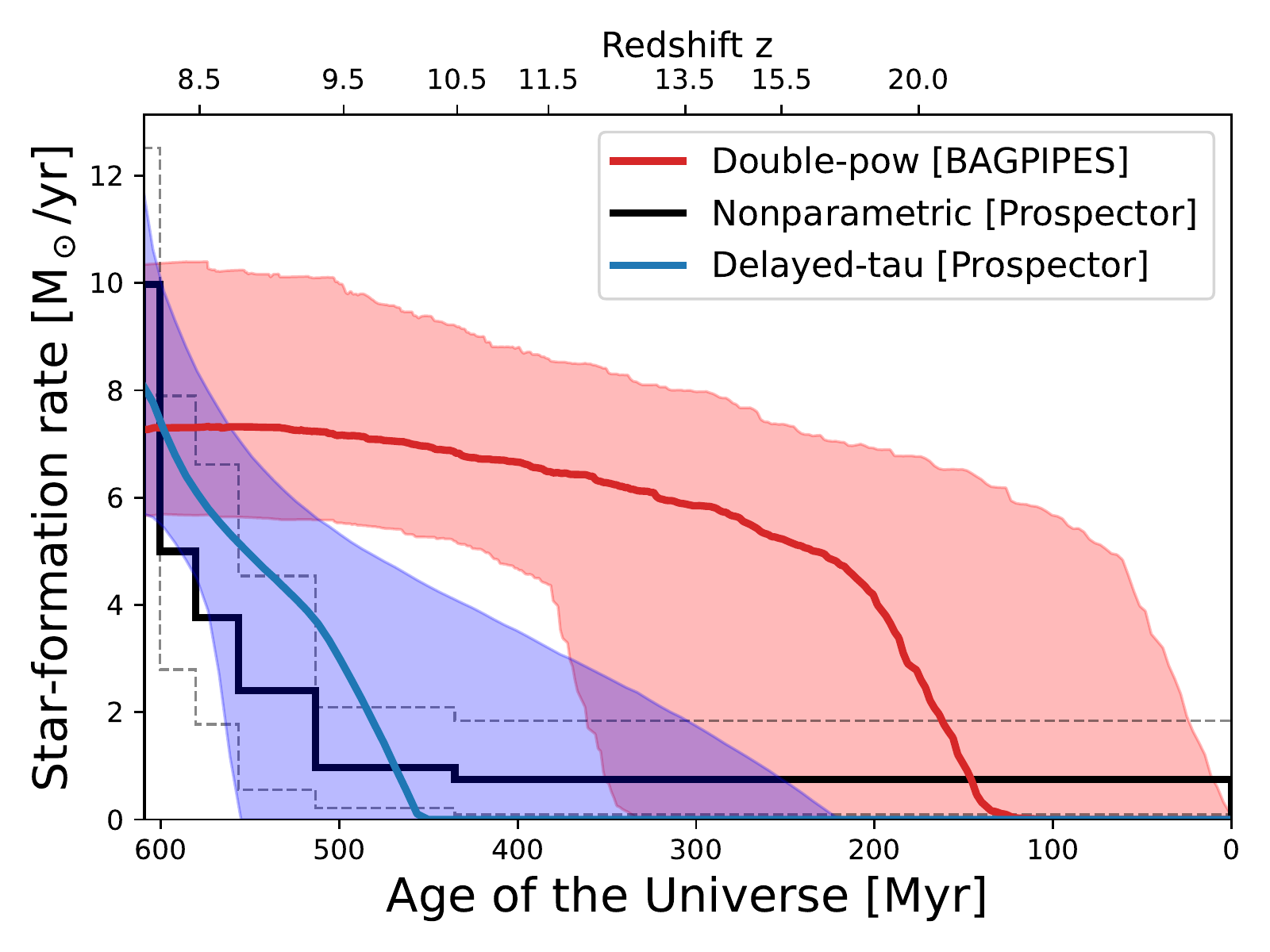} \\
 \vspace*{-1em}
 \caption{\small
 Star-formation history (SFH) of \galname, obtained from three independent methods.
 The SFH given by our default \bagp full spectrum fitting under the assumption of a double-power law model is represented by the red curve (best-fit value) with shaded regions (1-$\sigma$ confidence ranges).
 The SFH given by the \textsc{Prospector} SED fitting assuming a non-parametric and a delayed-$\tau$ models are shown in black histograms and blue curves, respectively.
 All three SFHs returned by different assumptions and techniques reach excellent consensus that the SFH of \galname is rapidly rising, in support of the existence of a very young stellar component likely associated with the Pop III signatures seen in the \jwst/NIRSpec spectra.
 }
 \label{fig:SFHs}
\end{figure*}

\bagp utilizes the nested sampling algorithm to perform efficient Bayesian inference and obtain posterior distribution of parameters. We take the [16th, 50th, 84th] percentiles of the parameter posteriors as the median and 1-$\sigma$ CI, and show the 1D posterior distributions of some key parameters in Fig.~\ref{fig:post}.
The overview of the physical properties of \galname is presented in Table~\ref{tab:gal}. 
After correcting for magnification ($\mu=2.26\pm0.14$), we obtain the following physical picture of \galname: it is a very low mass ($\Mstar\sim10^{7.8}\Msun$), young ($t_{\rm age}\sim216~{\rm Myrs}$) galaxy, actively forming stars (${\rm SFR \sim9.6 ~\Msun/yr}$) with sub-solar metallicity ($\log(Z_{\ast}/Z_{\odot})\sim-0.9$) and little dust ($A_{\rm V}\sim0.1$).
We recover a fast rising star-formation history of \galname, with a predominant young stellar population of 10-50 Myrs old.
The star-formation history (SFH) recovered by \bagp is shown in Fig.~\ref{fig:SFHs}, with the red curve and shaded region marking the best-fit and 1-$\sigma$ confidence range, respectively.

We follow the prescription in \citet{Alavi.2014} to estimate the absolute UV magnitude,
\begin{align}
    M_{\rm UV} = m_{\rm F140W} + \mu -5 \log(d_{\rm L}/{\rm 10~pc}) + 2.5 \log(1+z)
\end{align}
where $\mu$ is the lensing magnification factor expressed in magnitude units.
At $z=8.16$, the WFC3/F140W filter covers the rest-frame $\lambda_{\rm rest} \in[1300,1800]$ \AA and therefore appropriate to use.
The UV absolute magnitude is thus calculated to be {$M_{\rm UV}\sim-19.6~{\rm mag}$, which places \galname at the level of 40\% $L_*$ at $z\approx 8$ \citep{Bouwens.2015}.
Importantly, the bulk of the ionizing UV photons that caused the cosmic reionization are thought to come from sub-\Lstar systems \citep{Yan.2004c,Finkelstein.2019}.

\galname shows pronounced detection of the far UV continuum. We follow the standard formalism to derive its UV spectral slope ($\beta$).
We fit $f_{\lambda}\propto\lambda^{\beta}$ in the wavelength range of $\lambda_{\rm rest}\in[1300,2600]$ to the model galaxy SEDs produced by \bagp with all possible emission features properly masked.
The resulting UV slope is measured to be $\beta=-2.53_{-0.07}^{+0.06}$.
{We obtain consistent results if measuring $\beta$ using smoothed prism spectrum directly observed.}
This classifies \galname among the spectroscopically confirmed galaxies in the EoR having the steepest UV continuum slope --- a strong implication of significant leakage of its ionizing radiation to the IGM \citep{Bouwens.2010,Zackrisson.2013}.
The completion of reionization by $z\sim6$ requires that the \emph{absolute} escape fraction of the Lyman continuum (LyC) photons from galaxies should be $f^{\rm LyC}_{\rm esc}\gtrsim10\%$ on average \citep{Finkelstein.2012,Robertson.2015}.
In the EoR, a negative correlation between \Muv and $\beta$ has been seen from photometric analyses, in support of the dominant role played by the intrinsically faint systems in contributing to the IGM-ionizing photon budget \citep{Bouwens.2014,Bhatawdekar.2021}.
Using the HST/COS observations of the LyC signals from the Low-redshift Lyman Continuum Survey (LzLCS), \citet{Chisholm.2022} discovers that $\beta$ and the absolute escape fraction of Lyman continuum ($f^{\rm LyC}_{\rm esc}$) are strongly correlated:
\begin{align}\label{eq:beta_fesc}
    f^{\rm LyC}_{\rm esc} = (1.3\pm0.6)\times10^{-4}\cdot10^{(-1.22\pm0.1)\cdot\beta}.
\end{align}
We therefore obtain an estimate of $f^{\rm LyC}_{\rm esc}=0.16\pm0.03$ for \galname.

Fig.~\ref{fig:uv_slope} summarizes the currently available measurements of $\beta$ and $M_{\rm UV}$ of galaxies at $z\gtrsim7$, among which the spectroscopically confirmed sources are highlighted in color \citep{Watson.2015,Zitrin.2015t7c,Hashimoto.2018,Williams.2022,RB.2022,Morishita.2022,Schaerer.2022b,Curtis-Lake23,Bunker23}, and the photometric redshift (photo-$z$) selected candidates are represented in gray \citep{Tacchella.2022,Castellano.2022,Naidu.2022tc,Finkelstein.2022,Topping.2022,Cullen23a}. 
While steeper slopes ($\beta\sim-3$) have been reported for photo-$z$ and Lyman-break selected galaxies (see also \citet{Bouwens.2010}), \galname has by far the steepest UV spectral slope and thus the largest LyC escape fraction among all the spectroscopically confirmed galaxies at $z_{\rm spec}\geq7$.
Note that throughout this work, $\beta$ stands for $\beta_{\rm obs}$ in the nomenclature used by \citet{Chisholm.2022} without dust correction. After correcting for dust, \citet{Chisholm.2022} finds that the intrinsic UV spectral slopes ($\beta_{\rm int}$) of the LzLCS galaxies observed with \hst/COS are clustered in the range of [-2.8, -2.6].

\begin{figure*}
 \centering
 \includegraphics[width=\textwidth,trim=0cm 0cm 0cm 0cm,clip]{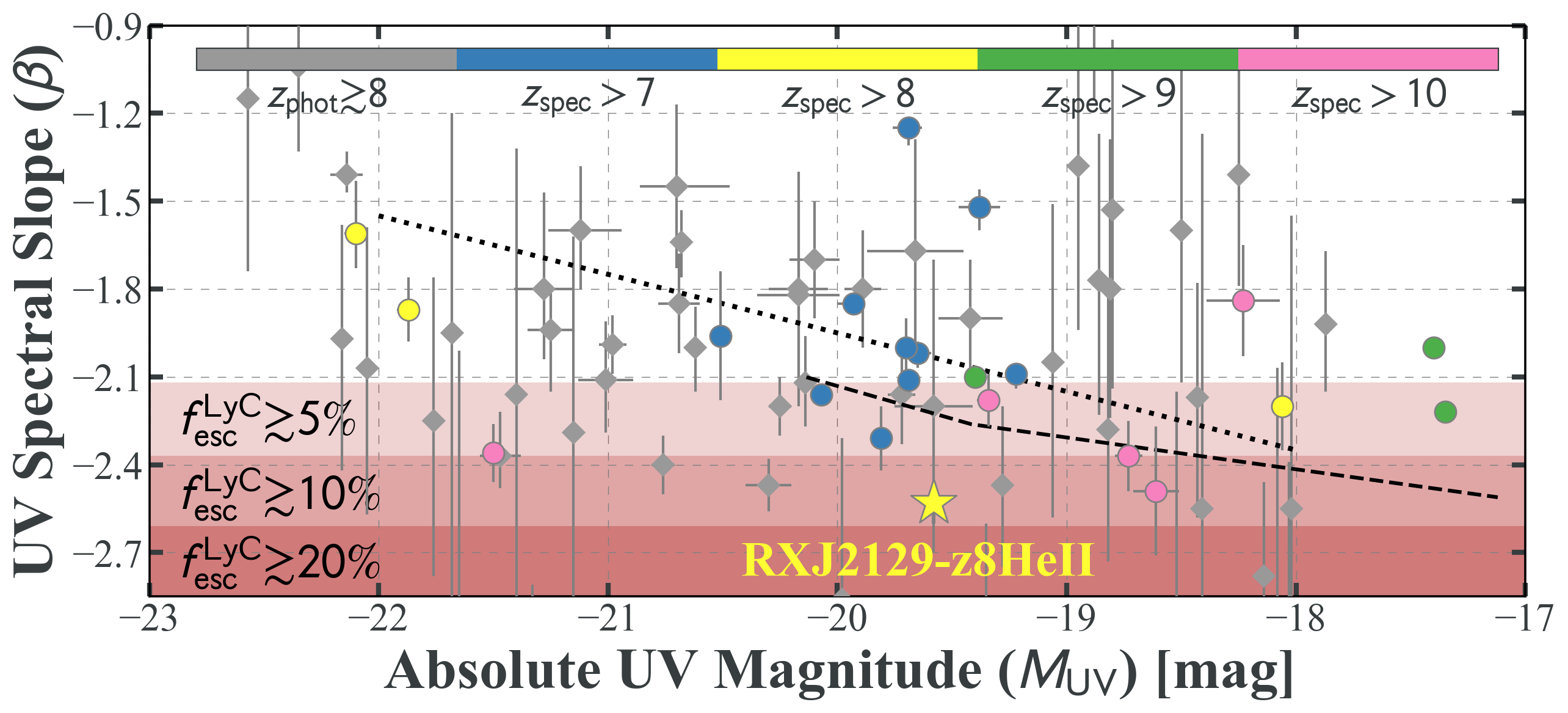} 
 \vspace*{-1em}
 \caption{\small
    UV spectral slope ($\beta$) as a function of absolute UV magnitude (\Muv). The color-coded circles represent the spectroscopically confirmed galaxies at $z_{\rm spec}>7$ with $\beta$ measurements \citep{Watson.2015,Zitrin.2015t7c,Hashimoto.2018,Williams.2022,Morishita.2022,Schaerer.2022b,Curtis-Lake23,Bunker23}. Among this cohort, \galname (indicated as the star) has the steepest slopes.
    The gray diamonds show the galaxy candidates with $z_{\rm phot}\gtrsim8$  \citep{Tacchella.2022,Castellano.2022,Naidu.2022tc,Finkelstein.2022,Topping.2022,Cullen23a}.
    The dotted and dashed lines denote the population-averaged correlation of \Muv and $\beta$ at $z\sim7$ \citep{Bouwens.2010,Bouwens.2014} and $z\sim8$ \citep{Bhatawdekar.2021}, respectively.
    The shaded regions correspond to the parameter space where the escape fraction of the ionizing radiation ($f^{\rm LyC}_{\rm esc}$) is above 5, 10, and 20 percent, respectively, derived by \citet{Chisholm.2022} from the LzLCS dataset. This strongly suggests that \galname is hitherto the most promising spectroscopically confirmed galaxy that enables significant LyC leakage into the IGM to contribute to the cosmic reionization.
 \label{fig:uv_slope}}
\end{figure*}

We also double check the derived stellar-population properties of \galname by modeling its broad-band photometry and spectra using the \textsc{Prospector} software \citep{Johnson.2021}. \textsc{Prospector} is built upon a fully Bayesian framework that makes it possible to fit the SED of high-redshift galaxies with complex SFH models. 
Regarding the basic fitting setup, we adopt the Flexible Stellar Population Synthesis (FSPS) code \citep{Conroy.2009,Conroy.2010} using the MIST stellar isochrone libraries \citep{Choi.2016,Dotter.2016} and the MILES stellar spectral libraries \cite{Falcon-Barroso.2011}. During the modeling, we use the MCMC sampling code dynesty \citep{Speagle.2020} based on the nested sampling technique. We assume the Kroupa IMF \citep{Kroupa.2001} for consistency with the IMF used in the nebular continuum from \citet{Byler.2017} and line emission model that we adopt in our SED modeling. We adopt the IGM absorption model from \citet{Madau.1995}. 
Following \citet{Charlot.2000}, we assume a two-component dust attenuation model where the dust attenuation of nebular emission and young stellar populations, and of old stellar populations, are treated differently.

We test our results with two different parameterizations of the galaxy’s SFH. First, we assume a piece-wise, nonparametric form composed of 6 lookback time bins (\ie, the time prior to the time of observation), where SFR is fitted as a constant in each bin. Among the 6 lookback time bins, the first two bins are fixed to be 0-5 and 5-10 Myr in order to capture recent star formation activities; the last bin is assumed to be $t(z=20) - t_{\rm H}$ where $t_{\rm H}$ is the Hubble time at $z=8.1623$; and the remaining 3 bins are evenly spaced in the logarithmic space of 10 Myr - $t(z=20)$. 
A similar setup of nonparametric SFH has been used in recent literature about high-redshift galaxy SED fittings \citep{Robertson23a,Tacchella23a}. During the nonparmetric SFH fitting, we adopt a continuity prior that has been demonstrated to work well across various galaxy types using mock observations from cosmological simulations \citep{Leja.2019n7o}.
The recovered SFH is represented by the black histograms shown in Fig.~\ref{fig:SFHs} with 1-$\sigma$ confidence region marked by the dashed histograms.
We note that the continuity prior is equivalent to a symmetric prior on stellar age and a constant SFH prior with ${\rm SFR}(t) = \Mstar/t_{\rm H}$. The fact that the fitting converges to a monotonically rising SFH suggests that the data strongly favor a very young stellar age of \galname, hence fully in line with its strong \HeII detection.

Second, we assume a parametric delayed-$\tau$ SFH model, \ie,
\begin{align}
    {\rm SFR}(t) \propto t\times{\rm e}^{-(t/\tau)}
\end{align}
with $\tau$ denoting the peak time of star formation sampled in the range of $\tau\in(0,t_{\rm H})$. The resulting SFH is shown as blue curve and shaded regions in Fig.~\ref{fig:SFHs} as well. Using two independent techniques (\ie \bagp and \textsc{Prospector}) under three sets of different SFH assumptions (\ie double power law, non-parametric, and delayed $\tau$ models), we achieve a self-consistent conclusion as follows. The SFH of \galname is sharply rising, and there exists a predominant stellar population 10-50 Myrs old.
This young stellar population is highly likely responsible for the strong \HeII line identified in the rest-frame UV spectra of \galname.

\subsection{Emission line fitting and diagnostics} \label{subsect:EL}

We utilize the \ppxf software \citep{Cappellari.2022} to perform accurate emission line analyses, fixing the source redshift to $z=8.1623$. 
To be self-consistent, we rely on the BC03 stellar population library \citep{Bruzual.2003}, generated using the Chabrier IMF \citep{Chabrier.2003} and dust corrected with the Calzetti reddening curve \citep{Calzetti.2000}.
The best-fit continuum model is shown as the red curve in the upper panel of Fig.~\ref{fig:fit}.
We then fit Gaussian profiles to nebular emission features at known wavelength positions, given fixed redshift.
During the emission line profile fitting, the range of their intrinsic width is set to [0, 300] km/s. The flux ratio of \OIII$\lambda\lambda$4959,5007 doublet is fixed to the theoretical value of 0.33. For the weak emission lines in the UV bands, we bind their shift of line centers to those stronger emission lines (e.g. \HeII) to improve the fitting robustness. We avoid fitting for any line emission blueward of $\lambda_{\rm rest}\sim1400$ \AA due to strong attenuation by the highly neutral IGM \citep{Dijkstra.2014}.

{\small
\begin{table*}
\centering
\begin{threeparttable}
\caption{Rest-frame UV and optical emission line properties of \galname.\label{tab:ELflux}}
\begin{tabular}{lcccc}
 \hline\hline 
 Emission Line  &  Observed Flux$^a$  & Intrinsic Flux$^b$     &  Rest-frame Equivalent Width \\
                & [$10^{-20}$\Funit]  & [$10^{-20}$\Funit] & [\AA] \\
 \hline \noalign {\smallskip}
    [O III]$\lambda$5008            &  779$\pm$21  &  390 $\pm$ 10  &  1015$\pm$83  \\ [\narrow] 
    H$\beta$                        &  141$\pm$20  &  71 $\pm$ 10  &  202$\pm$34  \\ [\narrow] 
    He II$\lambda$4686              &  <31         &  <16  &  <49  \\ [\narrow] 
    [O III]$\lambda$4363            &  <32         &  <16  &  <80  \\ [\narrow]
    H$\gamma$                       &  94$\pm$23   &  48 $\pm$ 12  &  203$\pm$69  \\ [\narrow]
    H$\delta$                       &  87$\pm$29   &  45 $\pm$ 15  &  139$\pm$50  \\ [\narrow] 
    [Ne III]$\lambda$3869           &  <43         &  <22  &  <44  \\ [\narrow] 
    [O II]$\lambda\lambda$3726,3729 &  64$\pm$23   &  33 $\pm$ 12  &  65$\pm$24  \\ [\narrow] 
    C III]$\lambda$1909             &  <70         &  <41  &  <13  \\ [\narrow]
    He II$\lambda$1640              &  206$\pm$38  & 120 $\pm$ 22   &  21$\pm$4  \\ [\narrow]
    C IV$\lambda$1549               &  <96         &  <57  &  <8  \\
 \hline \noalign {\smallskip}
 \end{tabular}
    \begin{tablenotes} 
      \item[NOTE ---] The observed line fluxes and equivalent widths are measured from our emission line analyses of the NIRSpec/MSA prism spectrum with the \ppxf software (see text for details). We list measurements for $\gtrsim2\sigma$ detections, or the 2-$\sigma$ upper limits. The error bars correspond to 1-$\sigma$ standard deviations. 
      \item[$a$] The observed line fluxes and upper limits before the corrections of dust extinction and lensing magnification.
      \item[$b$] The intrinsic line fluxes and upper limits after the corrections of dust extinction and lensing magnification ($\mu=2.26\pm0.14$). We adopt $A_{\rm V}=0.12\pm0.04$ measured from our spectro-photometric analyses to de-redden line fluxes since we only achieve upper limit on the nebular dust extinction.
    \end{tablenotes}
\end{threeparttable}
\end{table*}
}

The best-fit emission line profiles are demonstrated with orange curves in the upper panel of Fig.~\ref{fig:fit}. 
The resulting emission line fluxes (observed and intrinsic) and equivalent widths\footnote{EWs are always measured in rest frame throughout this paper.} (EWs) are listed in Table~\ref{tab:ELflux}.
Here the statistical uncertainties of the line fluxes and EWs are obtained from bootstrapping the fitting process to 50 mock spectra generated using the best-fit model spectrum and flux errors.

At a signal-to-noise ratio (SNR) threshold of $\gtrsim$3, we detect \Nline emission lines: $\OIII\lambda5008~(\defeq\OIII)$, \Hb, \Hg, \Hd, $\OII\lambda\lambda3727,3730~(\defeq\OII)$, and \HeII, amongst which \OIII, \Hb, and \HeII are detected with a SNR of 37, 7, 5, respectively.
In addition, we give 2-$\sigma$ upper limits for other lines in Table~\ref{tab:ELflux}, including \HeII$\lambda$4686, \OIII$\lambda$4363, \CIII, and \CIV.
We do not see any signature of \lya emission and put a 2-$\sigma$ upper limit of ${\rm EW}_{\lya}<10$ \AA.
The absence of \lya line is likely due to the damping wing opacity caused by the diffuse neutral IGM at $z\sim8$ and/or dense self-shielding H I gas clouds inside the large ionized H II bubbles \citep{Dijkstra.2014}.

\begin{figure}
 \centering
 \includegraphics[width=.48\textwidth,trim=0cm 0cm 0cm 0cm,clip]{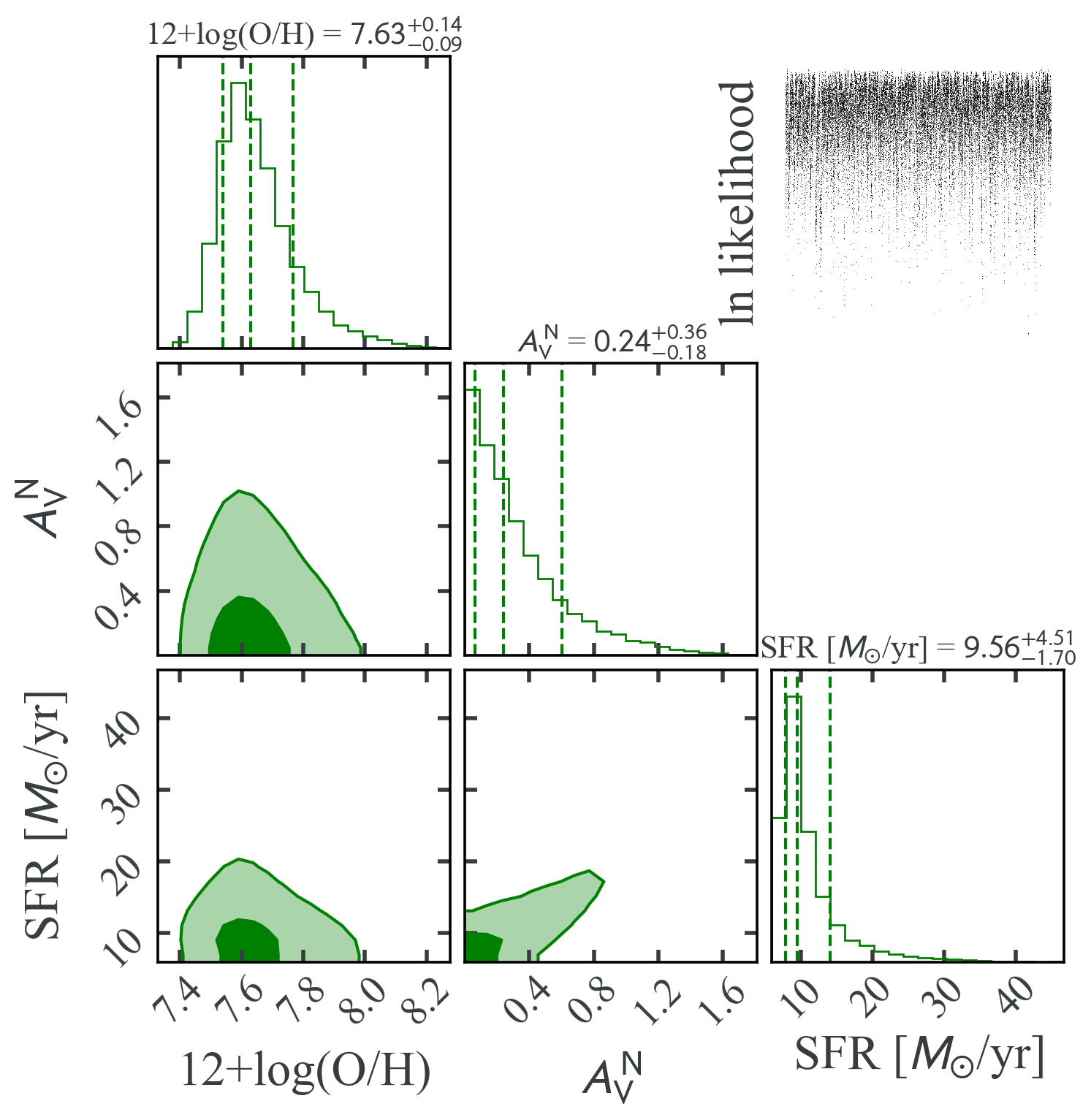}
 \vspace*{-1em}
 \caption{\small
 Corner plot showing the 1D/2D posterior constraints of the fitting parameters, \ie, gas-phase metallicity (\oh), nebular dust content ($A_{\rm V}^{\rm N}$), and instantaneous SFR already corrected for lensing magnification ($\mu=2.26\pm0.14$) of \galname, obtained from our Bayesian forward-modeling emission line diagnostic method.
 On top of each 1D posterior panel shown the resulting median with 1-$\sigma$ uncertainties for each parameter. Yet notice that we can only draw a 1-$\sigma$ upper limit on $A_{\rm V}^{\rm N}$ as presented in Table~\ref{tab:gal}.
 The panel in the upper right shows the $-\chi^2/2$ values for all the 80,000 parameter sampling iterations. They clearly reach a global minimum of $\chi^2$, implying a good convergence of our Bayesian inference.
 }
 \label{fig:oh12}
\end{figure}

We apply a well-established Bayesian forward-modeling inference framework to the measured line fluxes summarized in Table~\ref{tab:ELflux}. We constrain jointly three key properties of the interstellar medium (ISM): gas-phase metallicity (\oh), nebular dust extinction ($A^{\rm N}_{\rm V}$), and the instantaneous star-formation rate (SFR) converted from the de-reddened \Hb flux ($f_{\Hb}$) corrected for the magnification factor of $\mu=2.26\pm0.14$. Note that \oh and $A^{\rm N}_{\rm V}$ are unaffected by lensing magnification. The likelihood function is defined following \citet{Wang.2017,wangDiscoveryStronglyInverted2019,Wang.2020,Wang.2022a,Wang.2022b} as
\begin{align}\label{eq:chi2}
    \mathrm{L}\propto\exp\left(-\frac{1}{2}\cdot\sum_i \frac{\(f_{\el{i}} - R_i \cdot 
    f_{\Hb}\)^2}{\(\sigma_{\el{i}}\)^2 + \(f_{\Hb}\)^2\cdot\(\sigma_{R_i}\)^2}\right).
\end{align}
Here $f_{\el{i}}$ and $\sigma_{\el{i}}$ correspond to the intrinsic line (\OIII, \Hb, \Hg, \Hd, and \OII) flux and uncertainty with extinction corrected using the Cardelli dust attenuation law \citep{Cardelli.1989} with $R_{\rm V}=3.1$ following \citet{Valentino.2017}. $A^{\rm N}_{\rm V}\in(0, 3)$ is sampled as a free parameter. $R_i = f_{\el{i}}/f_{\Hb}$ and $\sigma_{R_i}$ represent the expected flux ratio and its intrinsic scatter of each line with respect to \Hb, given by the widely used Maiolino strong line calibrations \citep{Maiolino.2008} and the default Balmer decrements assuming case B recombination conditions. 
Thus the instantaneous SFR can be estimated from the Balmer line luminosity with the Kennicutt calibration \citep{Kennicutt.1998araa} and the Chabrier IMF \citep{Chabrier.2003}, assuming case B recombination:
\begin{align}\label{eq:kennicutt_sfr}
    {\rm SFR~[M_\odot {~\rm yr^{-1}}]}=1.3\times 10^{-41} \frac{L(\Hb) {\rm ~[erg ~s^{-1}]}}{\mu}.
\end{align}

We adopt the \emc software to perform the Markov Chain Monte Carlo Bayesian parameter sampling, with 100 of walkers each sampling 1000 iterations. After a burn-in of 200 for each walker, we have sampled the parameter space 80,000 times.
We show the resultant 1D and 2D parameter constraints in Fig.~\ref{fig:oh12}. The median constraints and 1-$\sigma$ CIs are shown in Table~\ref{tab:gal}.
To understand the energy source that powers the photoionization of the nebular emission observed in \galname, we rely on the mass-excitation diagram \citep[MEx][]{Juneau.2014,Coil.2015}, which is an empirical relation between \Mstar and the \OIII/\Hb flux ratio with demarcation schemes separating \HII region and AGN. 
We confirm non-detection of our galaxy in the archival Chandra data of in total $\sim$36k sec exposure (ObsID 9370, PI: Allen, and ObsID 553, PI Garmire) covering the galaxy cluster center field of RXJ2129.
Assuming a fiducial incident spectrum being an absorbed power law with a photon index of 1.7 and an absorption column density of $10^{21}$ cm$^{-2}$,
we reach a 3-$\sigma$ upper limit of $1.5\times10^{45}$ erg/s in the energy range of [0.5, 8] keV, lower than the measured X-ray luminosity of X-ray bright AGNs at $z>5$ \citep[see e.g.][]{liChandraDetectionThree2021}.
It is likely that \galname is a star-forming galaxy with negligible contamination from AGN ionization.
Although we caution that active AGNs need not be X-ray bright and the MEx diagram may differ substantially for very low metallicities.
The definitive exclusion of AGN contamination should come from much deeper \jwst NIRSpec spectroscopy with sufficient wavelength resolution to cast stringent limits on the high-ionization emission lines in the rest-frame UV and optical.

Since we only achieve upper limit on nebular dust extinction from our emission line diagnostics (see Fig.~\ref{fig:oh12}), we utilize $A_{\rm V}=0.12\pm0.04$ given by our spectro-photometric analyses in Sect.~\ref{subsect:bagp} to derive de-reddened line fluxes and uncertainties.
After correcting for dust attenuation, we obtain a high intrinsic flux ratio of $\OIII/\OII=11.7\pm4.2$, further supporting considerable LyC escape \citep{Izotov.2018}, consistent with our estimate based on $\beta$.
From our measurements of ${\rm EW}_{\OIII}$, we estimate the ionizing photon production efficiency to be $\log(\xion~{\rm [erg^{-1}Hz]})\sim25.55$ \citep{Tang.2019}, in agreement with the expectation that at $z\gtrsim8$ sub-\Lstar galaxies are the major sources of reionization \citep{Yan.2004c,Finkelstein.2019}.

Notably, the \HeII emission is the only rest-frame UV line clearly detected in \galname. This line has intrinsic (corrected for both lensing magnification and dust extinction) flux $f_{\HeII}=120\pm22~\times10^{-20}~\Funit$ and ${\rm EW_{\HeII}}=21\pm4$ \AA.
This is hitherto the highest redshift \HeII line potentially powered by star formation\footnote{
Recently, a tentative \HeII line detection (SNR$\sim$4) was reported by \citet{Bunker23} in GN-z11 at $z=10.603$. Unlike our galaxy, GN-z11 shows a plethora of emission features in its UV spectrum, including \CIV, \CIII, and $\NeIV\lambda\lambda2422,2424$, clear evidence of active galactic neucli (AGN) photoionization \citep{Maiolino.2023}.}.
The inset in the upper panel of Fig.~\ref{fig:fit} shows a zoom-in view of our fitting result to the source spectrum at $\lambda_{\rm rest}\in[1600, 1700]$, where \HeII and the oxygen auroral lines \ions{O}{iii}$\lambda\lambda$1661,1666 ($\defeq$\ions{O}{iii}) are marked in magenta vertical dashed lines. 
We verify that our wavelength calibration of the NIRSpec prism spectroscopy is sufficiently accurate with a 1-$\sigma$ uncertainty being $\sim$0.004 $\micron$ in the observer frame, corresponding to $\sim$4 \AA{} in the rest frame, distinguishing between \HeII and \ions{O}{iii} at a $\sim$5-$\sigma$ CI.
On the other hand, if the line were indeed \ions{O}{iii}, the only possible scenario that could power the doublets with such high EW$\sim$20 \AA{} is AGN ionization, which would result in much brighter \CIV and \CIII \citep{Hirschmann.2019}.
This reinforces our conclusion that the detection of the \HeII line is robust against negligible contamination from the oxygen auroral lines of \ions{O}{iii}.
We also perform detailed checks of the relative width and profile of the \HeII line, in comparison to those of \OIII-\Hb. We do not see any signs of broadening of \HeII, and find that the normalized profile and the relative width of \HeII match reasonably well those of \Hb and \OIII$\lambda$5007, strongly indicating that the \HeII line observed in \galname also originates from the nebular HII regions.

The nebular \HeII line requires a hard ionizing background radiation, which is usually attributed to Wolf-Rayet (WR) stars, stripped stars, X-ray binaries, or AGN \citep{Nanayakkara.2019,Saxena.2020}.
The location of \galname in the mass-excitation diagram \citep{Juneau.2014,Coil.2015}, the lack of variability from archival \hst imaging, and the non-detection in deep \textsc{Chandra} exposures altogether disfavor a supermassive black hole as the cause.
As opposed to \HeII, \galname shows absence of the UV carbon lines, setting it apart from other galaxies at various redshifts with \HeII detection \citep{Berg.2016,Sobral.2018cr7,Senchyna.2019cos,Nanayakkara.2019}, as shown in the right panel of Fig.~\ref{fig:HeII}.
Interestingly, some sources in the MUSE HUDF sample at $2<z<4$ \citep{Nanayakkara.2019} approach our measured limits of $f_{\HeII}/f_{\CIV}$ and $f_{\HeII}/f_{\CIII}$.
However, their \HeII EWs are reported to be less than 10 \AA\, not as high as that measured in our galaxy, and their \HeII emission is likely caused by stellar binarity and density-bounded HII regions with sub-solar ISM metallicities \citep{platConstraintsProductionEscape2019}.
We do not detect any UV metal lines (\eg \CIII, \CIV, \ions{O}{iii}, \ions{N}{iv}$\lambda\lambda$1483,1487, \ionp{N}{v}$\lambda$1240) either, inconsistent with the signatures of WR and stripped stars that are often seen in local galaxies \citep{Morris.2008,Leitherer.2020}.
Although we caution that the strength of these wind lines for WR and stripped stars are metallicity-dependent and is likely to be very weak at extremely low metallicities (e.g. much lower than one-tenth solar).

Alternatively, the \HeII line could be due to high-mass, metal-free population III (Pop III) stars that have exceedingly hard ionizing field capable of sufficient He$^+$ ionization.
While the ISM of \galname is already metal-enriched to roughly one-twelfth solar (\oh$\sim$7.63), its de-reddened flux ratio of $f_{\HeII}/f_{\Hb}=1.7\pm0.4$ is several orders of magnitude larger than that powered by ``normal'' metal-poor O/B stellar populations (see the left panel of Fig.~\ref{fig:HeII}).
We check that the UV and optical photometry of this potential ``Pop III'' component of our galaxy is consistent with the expected colors of Pop III galaxies reported in \citet{trusslerObservabilityIdentificationPopulation2023}.
In the next section, we investigate in detail the feasibility of reproducing the nebular emission observed in this system using Pop III stellar evolution and photoionization models.

\subsection{Pop III star photoionization model, formation rate, and total mass} \label{subsect:pop3_model}

\begin{figure*}
 \centering
 \includegraphics[width=.45\textwidth,trim=0cm 0cm 0cm 0cm,clip]{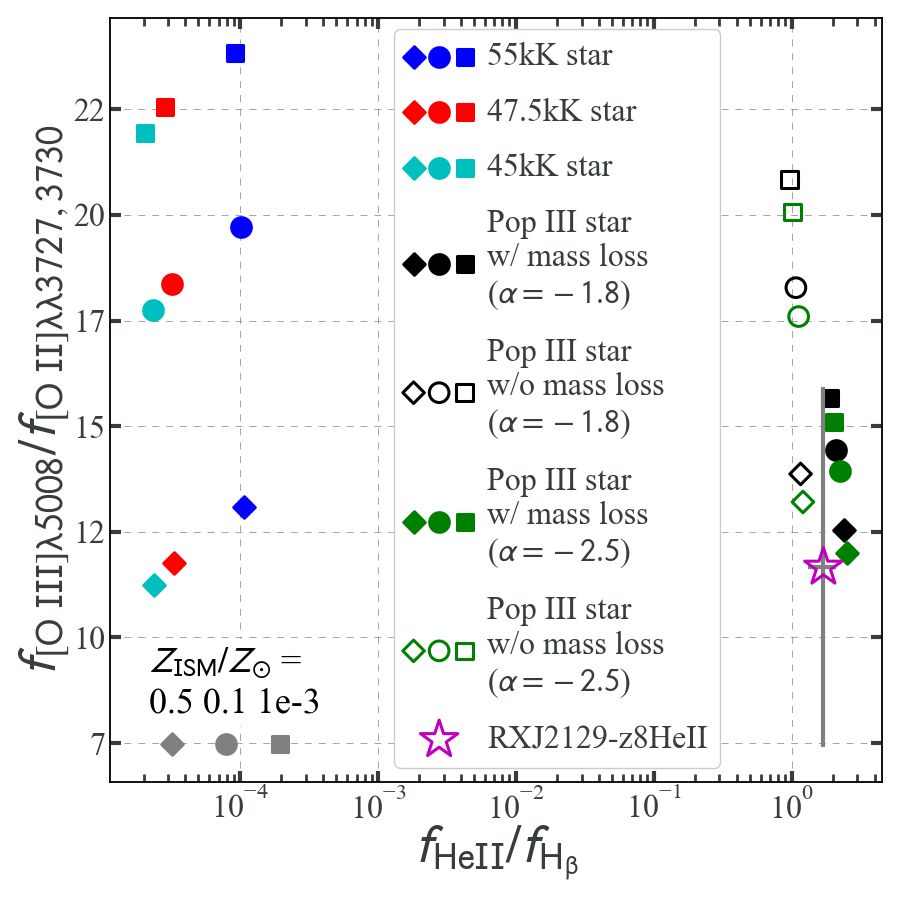}
 \includegraphics[width=.54\textwidth,trim=0cm 0cm 0cm 0cm,clip]{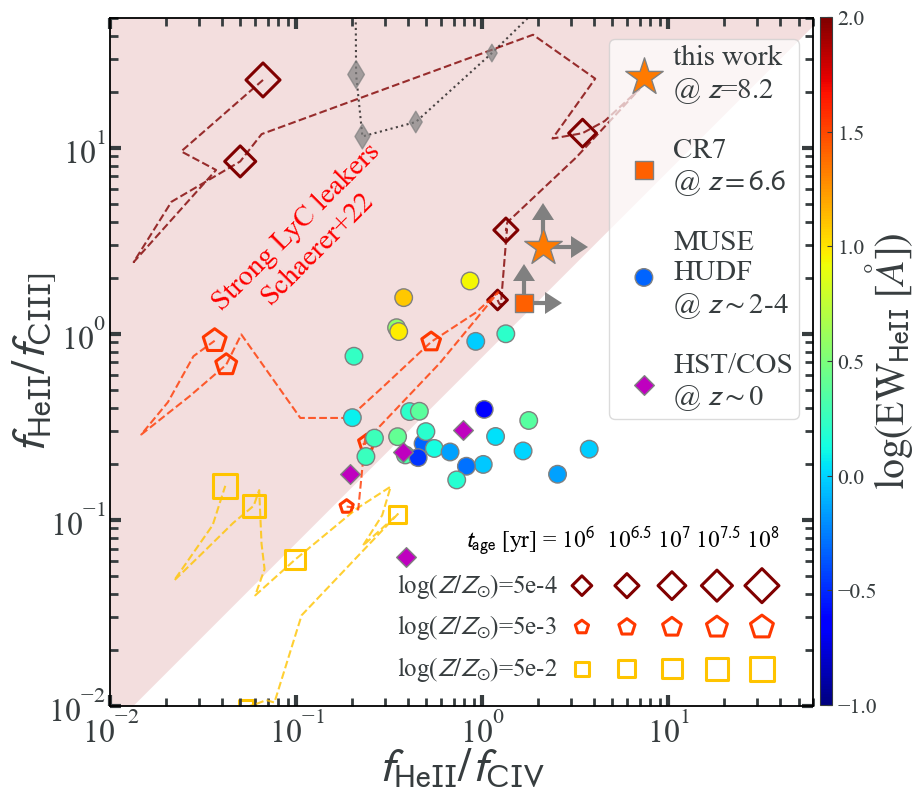}\\
 \caption{\small
    Emission line flux ratios from observations and models. 
    {\bf Left}: flux ratios of \OIII/\OII versus \HeII/\Hb predicted by the \textsc{MAPPINGS} V photoionization models \citep{Sutherland.1993} and measured in \galname (as in Table~\ref{tab:gal}) represented by the magenta star with 1-$\sigma$ error bars.
    The cyan, red, and blue symbols are calculated using the empirical spectra of metal-poor stars with atmospheric temperatures of 45kK, 47.5kK, and 55kK, respectively. The diamonds, circles and squares correspond to gas-phase metallicity ($Z_{\rm ISM}$) being half, one tenth, and one thousandth solar, respectively.
    It is obvious that ``normal'' O/B-type stellar populations cannot produce sufficient \HeII emission, in spite of their high \OIII/\OII at low $Z_{\rm ISM}$.
    We construct physically calibrated phenomenological models of Pop III star spectra, with different assumptions of mass loss and the power-law index ($\alpha$) in the Helium-ionizing energy range (see text). Encouragingly, all these models successfully reproduce the observed value of \HeII/\Hb. The mass loss models are better favored considering the observed value of \OIII/\OII.
    Crucially, we find that high \OIII/\OII \emph{alone} is not necessary a good indication of the hard ionizing spectra, potentially from Pop III stars.
    {\bf Right}: UV line ratios measured from galaxies that have reported \HeII detections (SNR$\gtrsim$2.5). 
    These measurements are color-coded in ${\rm EW}_{\HeII}$ reported in the respective work. The arrows denote 2-$\sigma$ lower limits. \galname shows greatly elevated \HeII/\CIII and \HeII/\CIV flux ratios, as compared to local galaxies \citep{Berg.2016,Senchyna.2019cos}, galaxies at the cosmic noon epoch \citep{Nanayakkara.2019}, and CR7 at $z=6.6$ \citep{Sobral.2018cr7}.
    We also overlay three sets of evolutionary tracks at different stellar metallicities given by the BPASSv2.1 stellar population synthesis models \citep{Xiao.2018}.
    Among these three tracks, the measured flux ratio lower limits of \galname clearly prefer that with extremely low metallicity.
    The grey diamonds represent the predictions of AGN narrow-line regions calculated using the CLOUDY code \citep{ferlandCLOUDY90Numerical1998}, with
    larger symbol size corresponding to higher metallicity (i.e. 0.03, 0.1, 0.5 and 1$\times$ solar).
 }
 \label{fig:HeII}
\end{figure*}

One crucial aspect of this work is to explain the flux ratios measured between the strong emission lines observed in the rest-frame UV and optical spectra of \galname, using state-of-the-art photoinization models and reasonable assumptions on the input ionizing spectra from Pop III stellar populations. We carry out such analysis with the state-of-the-art \textsc{MAPPINGS} V photoionization code \citep{Sutherland.1993}. \textsc{MAPPINGS} V has comprehensive consideration of microphysics in ISM, providing accurate predictions of fluxes of over 80,000 cooling and recombination lines.
By propagating a simple toy model of Pop III star ionizing spectra calibrated by \citet{Schaerer.2002,Schaerer.2003} through the \textsc{MAPPINGS} V photoionization code, we verify that the observed line flux ratios can be reproduced under reasonable assumptions of the ionizing spectral hardness and spectral shapes that are believed to be typical in Pop III stellar populations.
We assume the following form of broken power laws for this toy model:
\begin{subnumcases}{f_\nu=\label{eq:popIII_models}}
   C_0 \cdot \nu^{-\beta} \quad   h\nu<13.6~{\rm eV}   \label{C0},     \\
   C_1 \cdot \nu^{-1}  \quad   h\nu\in[13.6, 24.6)~{\rm eV}   \label{C1},    \\
   C_2 \cdot \nu^{\alpha}  \quad   h\nu\in[24.6, 54.4)~{\rm eV}   \label{C2},  \\
   C_3 \cdot \nu^{\alpha}  \quad   h\nu\geq 54.4~{\rm eV}   \label{C3},
\end{subnumcases}
where $\beta=-2.53$ is determined from our full spectrum fitting analysis and $\alpha$ represents the power-law index in the He$^0$ and He$^+$ ionization energy range. 
We experiment with two indices of $\alpha=-1.8$ and $\alpha=-2.5$ in constructing this phenomenological model of Pop III star EUV spectra, and arrive at a conclusion that $\alpha$ does not affect the resulting line ratios as long as the spectral hardness is maintained (see Table~\ref{tab:model_line_ratios}). 
The $C_0$ coefficient is computed to match the measured rest-frame equivalent width of ${\rm EW_{\HeII}}=21$ \AA, assuming that observationally the far UV continuum is mostly coming from Pop III stars \footnote{$C_0$ sets the continuum level of the non-ionizing spectrum (\ie Eq.~\ref{C0}) at the rest-frame wavelength of 1640 \AA, where the \HeII line resides. Our simple but physically calibrated phenomenological models of the Pop III ionizing spectrum (Eqs.~\ref{C1}-\ref{C3}) can predict the total line flux of the \HeII line. The value of $C_0$ can therefore be calculated using this line flux and the observed equivalent width of \HeII in \galname.}. 
The other coefficients (\ie $C_{1,2,3}$) are determined to match the hardness ratios of the ionizing spectra reported in \citet{Schaerer.2002}, given by the ratios of the ionizing photon flux ($Q({\rm H})$, $Q({\rm He^0})$, and $Q({\rm He^+})$).
We take the time-averaged $Q$ values calculated from the Pop III stellar evolution tracks with $M_{\rm ini}=500 \Msun$ as specified in Table 4 and 5 of \citet{Schaerer.2002}, to account for the effects of stellar evolution which typically decreases the spectral hardness by a factor of 2. 
Conventionally, Pop III stars with very high mass are preferred because of the lack of efficient cooling channels (due to the absence of the CNO elements; see reviews by \citet{Bromm.2004,Bromm.2013}).
Both scenarios with and without mass loss are considered to investigate the effects of mass loss due to stellar winds and core-collapsed supernovae on ratios of the emergent rest-frame UV/optical emission lines \citep{Schaerer.2003}.
The resulting phenomenological models of Pop III star ionizing spectra, calibrated against realistic Pop III stellar evolution tracks and atmosphere models are represented by the black and green lines in Fig.~\ref{fig:popIII_models}.
For comparison, we also calculate the empirical ionizing spectra of single metal-poor stars using the \textsc{TLUSTY} non-LTE plane parallel code modeling massive star atmosphere \citep{Hubeny.2011,Hubeny.2021}, with three characteristic temperatures of 55kK, 47.5kK and 45kK, shown as the blue, red, and cyan curves, respectively, in Fig.~\ref{fig:popIII_models}.
Their stellar-phase metallicity is taken to be [Fe/H]=-3 (one thousandth solar).
We assume three conditions of the gas-phase metallicity for all input stellar spectra: $Z_{\rm ISM}/Z_{\odot}=0.5, 0.1, 10^{-3}$ in our photoionization calculations.

{\small
\begin{table*}
\centering 
\begin{tabular}{llcccc} 
\hline\hline 
 Star Type  &  Spectral Model  &  \OIII/\Hb  & \OII/\Hb & \HeII/\Hb & \OIII/\OII \\
\hline
 O/B  &  55.0kK, 0.5$Z_\odot$  &  13  &  0.99  &  0.00011   &  13 \\ [\narrow] 
 O/B  &  55.0kK, 0.1$Z_\odot$  &  5.7  &  0.29  &  0.0001   &  20 \\ [\narrow] 
 O/B  &  55.0kK, 0.001$Z_\odot$  &  0.073  &  0.0031  &  9.1e-05   &  24 \\ [\narrow] 
 O/B  &  47.5kK, 0.5$Z_\odot$  &  11  &  0.95  &  3.3e-05   &  12 \\ [\narrow] 
 O/B  &  47.5kK, 0.1$Z_\odot$  &  5.5  &  0.3  &  3.2e-05   &  18 \\ [\narrow] 
 O/B  &  47.5kK, 0.001$Z_\odot$  &  0.074  &  0.0033  &  2.9e-05   &  23 \\ [\narrow] 
 O/B  &  45.0kK, 0.5$Z_\odot$  &  11  &  0.94  &  2.4e-05   &  11 \\ [\narrow] 
 O/B  &  45.0kK, 0.1$Z_\odot$  &  5.4  &  0.3  &  2.3e-05   &  18 \\ [\narrow] 
 O/B  &  45.0kK, 0.001$Z_\odot$  &  0.074  &  0.0034  &  2.1e-05   &  22 \\ [\narrow] 
 Pop III & mass loss, $\alpha$=-1.8, 0.5$Z_\odot$  &  14  &  1.1  &  2.4   &  13 \\ [\narrow] 
 Pop III & mass loss, $\alpha$=-1.8, 0.1$Z_\odot$  &  3.7  &  0.26  &  2.1   &  14 \\ [\narrow] 
 Pop III & mass loss, $\alpha$=-1.8, 0.001$Z_\odot$  &  0.039  &  0.0025  &  1.9   &  16 \\ [\narrow] 
 Pop III & no mass loss, $\alpha$=-1.8, 0.5$Z_\odot$  &  15  &  1.1  &  1.2   &  14 \\ [\narrow] 
 Pop III & no mass loss, $\alpha$=-1.8, 0.1$Z_\odot$  &  5.2  &  0.28  &  1.1   &  18 \\ [\narrow] 
 Pop III & no mass loss, $\alpha$=-1.8, 0.001$Z_\odot$  &  0.059  &  0.0028  &  0.96   &  21 \\ [\narrow] 
 Pop III & mass loss, $\alpha$=-2.5, 0.5$Z_\odot$  &  13  &  1.1  &  2.5   &  12 \\ [\narrow] 
 Pop III & mass loss, $\alpha$=-2.5, 0.1$Z_\odot$  &  3.7  &  0.26  &  2.2   &  14 \\ [\narrow] 
 Pop III & mass loss, $\alpha$=-2.5, 0.001$Z_\odot$  &  0.039  &  0.0026  &  2   &  15 \\ [\narrow] 
 Pop III & no mass loss, $\alpha$=-2.5, 0.5$Z_\odot$  &  14  &  1.1  &  1.2   &  13 \\ [\narrow] 
 Pop III & no mass loss, $\alpha$=-2.5, 0.1$Z_\odot$  &  5.1  &  0.29  &  1.1   &  18 \\ [\narrow] 
 Pop III & no mass loss, $\alpha$=-2.5, 0.001$Z_\odot$  &  0.058  &  0.0029  &  1   &  20 \\
 \hline
 Observation  &  \galname  & 5.5$\pm$0.8 & 0.5$\pm$0.2 & 1.7$\pm$0.4 & 11.7$\pm$4.3 \\
 \hline
\end{tabular}
\caption{Emission line flux ratios calculated by our \textsc{MAPPINGS} V photoionization modeling framework, assuming empirical and phenomenological stellar spectra of normal O/B and Pop III stars.
The last row shows our actual measurements with 1-$\sigma$ uncertainties of these key line ratio diagnostics in \galname using \jwst/NIRSpec prism spectroscopy.
\label{tab:model_line_ratios}}
\end{table*}
}

The \textsc{MAPPINGS} V photoionization code takes these ionizing spectra as input and produce the emergent rest-frame UV and optical spectra via self-consistent Monte Carlo radiative transfer. We adopt the plane-parallel geometric configuration in radiative transfer calculations, since we believe that the main ionizing sources (Pop III stars) are located $\sim$1 kpc away (\ie in component B) from the ISM where our NIRSpec slit spectroscopy is taken (\ie in component A).
The resulting line flux ratios of \OIII/\Hb, \OII/\Hb, \HeII/\Hb and \OIII/\OII from these empirical and phenomenological stellar spectral models are given in Table~\ref{tab:model_line_ratios}.
The left panel of Fig.~\ref{fig:HeII} shows the model predictions of \OIII/\OII versus \HeII/\Hb, compared against our observations. 
Our detailed photoionization modeling clearly suggests that normal O/B stellar populations, albeit with high atmospheric temperatures and high \OIII/\OII, do not suffice to power prominent high-ionization \HeII emission, as they would only result in insignificant \HeII/\Hb ratio lower than our observation by 4-5 orders of magnitude. 
This strongly indicates that although \OIII/\OII is widely used as a key diagnostic of the ionization state of the gas in extragalactic H II regions \citep{Kewley.2013}, it is not necessarily a good proxy \emph{by itself} for identifying potential Pop III stars in metal-poor systems, due to serious contamination from normal O/B stellar populations.

The Pop III star scenarios, on the other hand, offer sufficient extreme UV ionization field giving rise to strong \HeII emission and can reproduce the observed ratio of \HeII/\Hb=$1.7\pm0.4$ in \galname.
We find that both the Pop III ionizing spectra with and without mass loss lead to comparable ionizing radiation, with the former producing slightly elevated \HeII/\Hb ratios, due to increased spectral hardness \citep{Schaerer.2002,Schaerer.2003}.
The observed ratio of \OIII/\OII=$11.7\pm4.2$ better favors the Pop III spectral models with strong mass loss.
Finally, we also see good agreement between the observed \OIII/\Hb=$5.5\pm0.8$ and our calculations setting ISM metallicity to one tenth solar (see Table~\ref{tab:model_line_ratios} and circles in the left panel of Fig.~\ref{fig:HeII}), which lends support to the constraint on \oh from our emission line diagnostics shown in Fig.~\ref{fig:oh12}.

\begin{figure*}
 \centering
 \includegraphics[width=.8\textwidth,trim=0cm 0cm 0cm 0cm,clip]{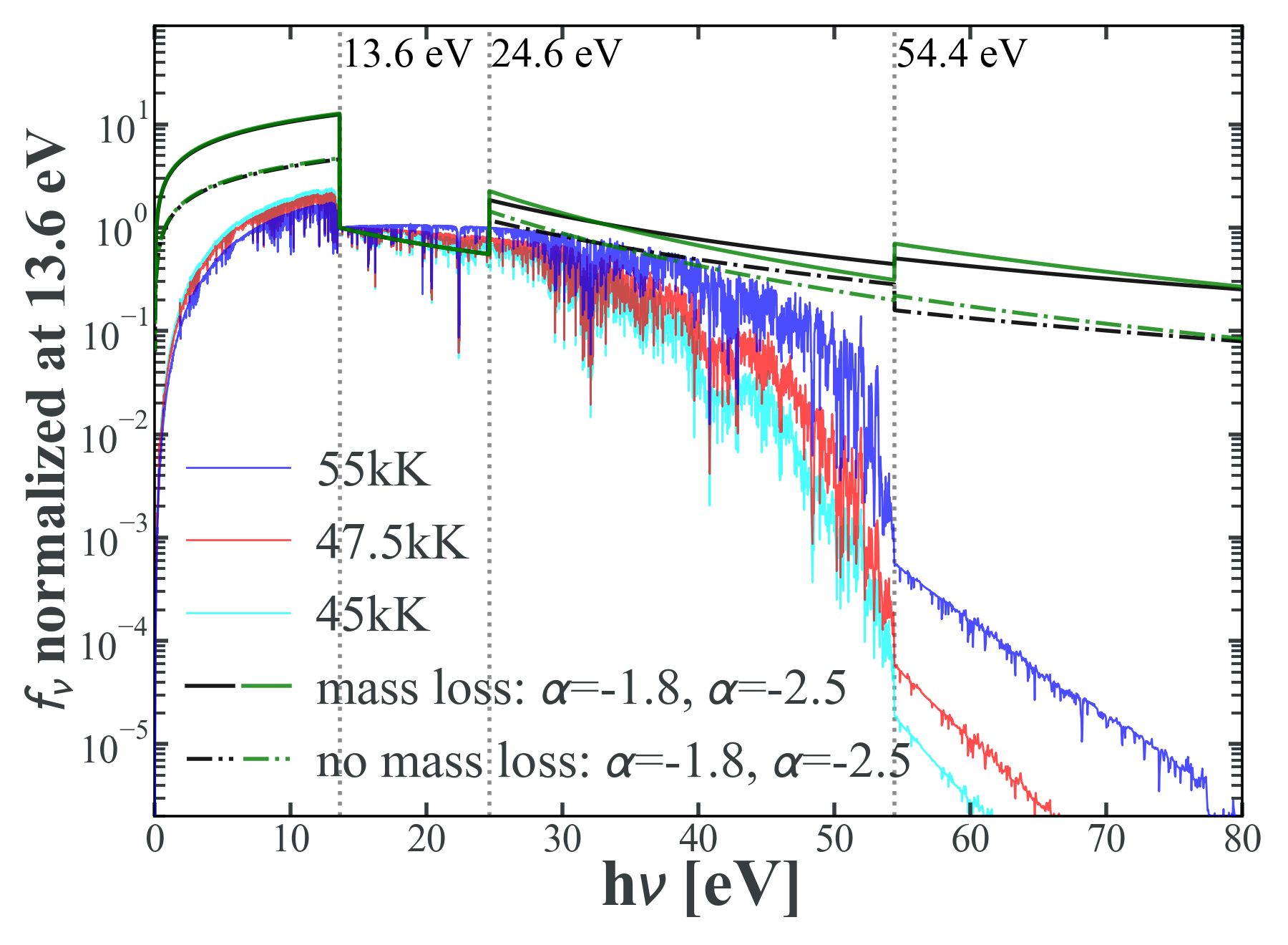}
 \vspace*{-1em}
 \caption{\small
 Realistic stellar spectra input to our photoionization modeling framework.
 The cyan, red, and blue curves correspond to the empirical spectra of metal-poor ([Fe/H]=-3) single stars with high atmospheric temperatures of 45kK, 47.5kK, and 55kK, respectively, representing the typical populations of low-metallicity O/B stars.
 These empirical star model atmospheres are constructed using the non-LTE plane-parallel code \textsc{TLUSTY} \citep{Hubeny.2011,Hubeny.2021}.
 The black and green lines show our broken power law models of Pop III stellar spectra, defined in Eq.~\ref{eq:popIII_models}, with different assumptions of mass loss and the power-law index $\alpha$ (see text for more details).
 The ionizing sections of these Pop III spectra are calibrated by the time-averaged hardness ratios --- described by $Q({\rm He^0})/Q({\rm H})$ and $Q({\rm He^+})/Q({\rm H})$ tabulated in \citet{Schaerer.2002} --- with and without mass loss.
 The non-ionizing sections are determined based upon our actual measurement of the UV spectral slope ($\beta=-2.53$) and the emission line equivalent widths (\ie ${\rm EW_{\HeII}}=21$ \AA).
 The three vertical dotted gray lines mark the energy levels of 13.6, 24.6, and 54.4 eV, corresponding to the ionization of H , He$^0$, and He$^+$ ions.
 \label{fig:popIII_models}}
\end{figure*}

To estimate SFR of Pop III star (${\rm SFR_{Pop III}}$) and compare with the instantaneous SFR converted from the de-reddened \Hb line luminosity (${\rm SFR_{O/B}}=9.56_{-1.70}^{+4.51}$ as shown in Table~\ref{tab:gal}), we follow the prescriptions of \citet{Raiter.2010,caiCONSTRAININGVERYHIGH2015}. Under the fiducial conditions of case B recombination and ionization-bounded nebula with constant density and temperature, the \HeII line luminosity generated by Pop III stars can be expressed as a function of He$^+$ ionizing photon flux ($Q({\rm He^+})$) and Pop III SFR, \ie,
\begin{align}\label{eq:popIII_SFR}
    L_{\HeII} & = c_{1640}(1-\fesc)Q({\rm He^+})\left(\frac{\rm SFR_{Pop III}}{[\Msun~{\rm yr}^{-1}]}\right)    \nonumber \\
        & = f_{1640} (1-\fesc) \left(\frac{\rm SFR_{Pop III}}{[\Msun~{\rm yr}^{-1}]}\right),
\end{align}
where $f_{1640}$ is the theoretical \HeII line luminosity for constant star formation models normalized to ${\rm SFR} = 1~\Msun~{\rm yr}^{-1}$, given in Table~7 of \citet{Schaerer.2002} (also see Table~\ref{tab:popIII_SFR}). 
From our emission line analyses, we measured the \HeII line flux corrected for both lensing magnification and dust extinction to be $f_{\HeII}=120\pm22\times10^{-20}~\Funit$, corresponding to a \HeII luminosity of $L_{\HeII} = 9.7\pm1.8\times10^{41}~\Lunit$.
Therefore, we can estimate ${\rm SFR_{Pop III}}$ under various Pop III stellar evolution models as shown in Table~\ref{tab:popIII_SFR}, from the observed $L_{\HeII}$ and the escape fraction of $\fesc=16\%$ inferred from the UV spectral slope of $\beta=-2.53$.
We caution that there is a large uncertainty attached to the conversion from $\beta$ to $\fesc$, and the escape fractions of the ionizing flux in different energy regimes (e.g. \fesc($>$13.6 eV) vs. \fesc($>$54.4 eV)) can be quite different.
In fact, the galaxies for which the $\beta$-$\fesc$ relation has been derived empirically (Eq.~\ref{eq:beta_fesc}) are not systems that host Pop III stars \citep{Chisholm.2022}.
Applying this relation to potential Pop III star host galaxies might be expected to under-estimate $\fesc$, since the Pop III stellar models with extremely top-heavy initial mass functions predict relatively red UV slopes due to increased nebular continuum \citep{cameronNebularDominatedGalaxies2023}.
Nevertheless, with discretion, we estimate the ratio between ${\rm SFR_{Pop III}}$ and $\rm SFR_{O/B}$ calculated using Eq.~\ref{eq:kennicutt_sfr} to be in the range of 3--72\%.
Assuming the Eddington limit and $L_{\HeII}$ being $\sim$1\% of the total bolometric luminosity, we can also derive the total mass of the Pop III stars to be $7.8\pm1.4\times10^5\Msun$, consistent with the predictions based on numerical simulations \citep{Venditti.2023}.

{\small
\begin{table*}
\centering 
\begin{tabular}{cccccc} 
\hline\hline 
 Model ID &  IMF mass range & Mass Loss & $f_{1640}$ & $\rm SFR_{Pop III}$ & $\frac{\rm SFR_{Pop III}}{\rm SFR_{O/B}}$ \\
 in \citet{Schaerer.2002}  &  $M/\Msun$  &  & [$\mbox{ergs}\ \rm{s}^{-1}$] & $[\rm M_\odot ~yr^{-1}]$ & \\
\hline
 B & [1,500] & No    & $9.98\times10^{40}$ & 6.86  & 0.72   \\
 C & [50,500] & No   & $8.38\times10^{41}$ & 0.82  & 0.08   \\
 D & [1,500] & Yes   & $3.12\times10^{41}$ & 2.19  & 0.23   \\
 E & [50,1000] & Yes & $2.33\times10^{42}$ & 0.29  & 0.03   \\
\hline
\end{tabular}
\caption{Pop III Star Formation Rate (SFR) of \galname at $z=8.1623$ inferred from Eq.~\ref{eq:popIII_SFR}.
Note that $f_{1640}$ stands for the theoretical \HeII luminosity normalized to ${\rm SFR} = 1~\Msun~{\rm yr}^{-1}$ for the four stellar evolution models \citep{Schaerer.2002}. The actual observed \HeII luminosity of \galname is $L_{\HeII} = 9.7\pm1.8\times10^{41}~\Lunit$. The last column represents the ratio between the Pop III SFR ($\rm SFR_{Pop III}$) and the instantaneous SFR measured from the intrinsic \Hb luminosity using Eq.~\ref{eq:kennicutt_sfr} ($\rm SFR_{O/B}$). Note that the Kennicutt calibration \citep{Kennicutt.1998araa} used here to calculate $\rm SFR_{O/B}$ is generally applicable to galaxies with strong recombination lines powered by the ionizing flux predominantly from O/B-type stars who typically have $\Mstar>10~\Msun$ and life times $t_{\rm age}<20 ~{\rm Myr}$.
Here $\rm SFR_{Pop III}$ has been converted to the value compatible with the Chabrier IMF \citep{Chabrier.2003}, to facilitate a direct comparison with $\rm SFR_{O/B}$.
\label{tab:popIII_SFR}}
\end{table*}
}

It is encouraging to see that the SFR ratio ($\frac{\rm SFR_{Pop III}}{\rm SFR_{O/B}}$) and the total mass of Pop III stars within this system reside in reasonable ranges. 
Nevertheless we caution the readers about the caveat that some assumptions are still highly uncertain and demand more work in this fast rising field. First and foremost, the theoretical framework of \citet{Schaerer.2002,Schaerer.2003,Raiter.2010} is based upon the top-heavy Salpeter \citep{Salpeter.1955}\ IMF\footnote{We have applied the conversion between the Salpeter and Chabrier IMFs when deriving ${\rm SFR_{Pop III}}$, presented in Table~\ref{tab:popIII_SFR}.}. A factor of eight difference in the inferred SFR is caused by different low mass cutoffs in IMF mass range. 
Secondly, we see that models incorporating strong mass loss decrease SFR by a factor of three. However, the detailed physical mechanisms (\eg rotation, pair-instability supernovae, multiplicity, etc.) driving such high mass loss are still under investigation \citep{Ekstrom.2008}.
Lastly, we caution that the \HeII line emission coefficient $c_{1640}=5.67\times10^{-12}~{\rm erg}$ is computed for case B with nebular electron density $n_{\rm e}=100~{\rm cm}^{-3}$ and $T_{\rm e}=30~{\rm kK}$, under the assumption that the radiation field is optically thick. Realistically, there can exist steep density/temperature gradients and significant deviations from the fiducial case B conditions around extremely hot Pop III stars.
In particular, the relatively high LyC escape fraction ($\fesc^{\rm LyC}=16\%$) estimated in \galname, indicated by its large O32 flux ratio ($f_{\OIII}/f_{\OII}=11.7$) and blue UV continuum slope ($\beta=-2.53$), is in favor of the ionization-bounded HII regions, rather than the density-bounded ones assumed in case B. This density-bounded condition can suppress low-ionization emission lines such as \Hb and potentially artificially boost ratios between high-ionization and low-ionization lines such as \HeII/\Hb \citep{platConstraintsProductionEscape2019}.
Nonetheless, our work presents the critical first step forward to apply physically motivated Pop III stellar evolution models to \jwst high-quality NIRSpec spectroscopy of galaxy candidate revealing Pop III star signatures.

\section{Conclusions and discussion}\label{sect:conclu}

In this paper, through the comprehensive analysis of the novel \jwst NIRSpec prism spectroscopy and NIRCam imaging, we present \galname --- an intriguing galaxy in the EoR. 
Fitting simultaneously the entire spectroscopic (continuum + lines) and the photometric data sets, we confirm \galname as an extreme emission line galaxy with strong continuum emission at $z_{\rm spec}=8.1623\pm0.0007$.
It has a sub-\Lstar $M_{\rm UV}$, high-EW rest-frame optical lines, and the steepest UV spectral slope ($\beta=-2.53_{-0.07}^{+0.06}$) amongst all spectroscopically confirmed galaxies at $z\gtrsim7$ reported to date, implying that it has a large LyC escape fraction $f^{\rm LyC}_{\rm esc}\sim16\%$, the largest among its cohort. Therefore, \galname is a promising representative of the predominant galaxy populations that are capable of producing and \emph{leaking} their extreme UV photons to the IGM, causing the bulk of the neutral hydrogen in the IGM ionized.
Furthermore, its prominent \HeII line and the large flux ratios between \HeII and UV metal and Balmer lines make it the best candidate to date where this very galaxy could have a significant Pop III stellar constituent within its stellar populations.

Using the \textsc{MAPPINGS} V photoionization code and a series of phenomenological models of Pop III star spectra calibrated against theoretical Pop III stellar evolution models, we can successfully reproduce the measured ratios of \HeII/\Hb and \OIII/\OII, assuming strong mass loss and one-tenth ISM metallicity.
Crucially, we also find that high \OIII/\OII \emph{alone} is not a good indication of the presence of Pop III stars, providing important guidelines for future \jwst surveys with the aim of identifying Pop III signatures in metal-poor systems in the EoR.
All this, together with the large ${\rm EW_{\HeII}}\sim21$ \AA, suggests that \galname likely has a mixture of metal-poor O/B and Pop III stars, the latter of which are the energy source of the \HeII line emission.
With the caveat that our \fesc estimate based on observed UV spectral slope can be an underestimation due to nebular continuum, we estimate the ratio between the formation rates of Pop III and O/B stars in the range of [3, 72]\%, and derive the total mass of Pop III stars within \galname to be $7.8\pm1.4\times10^5\Msun$, consistent with the predictions based on numerical simulations.

The putative Pop III stars could either be spatially mixed with the enriched populations or segregated. Indeed, recently cosmological hydrodynamic simulations suggest that Pop III stars can continue to form till the end of the EoR, triggered by sustained accretion of pristine gas from the cosmic web onto galaxy disks, resulting in ``hybrid'' Pop III galaxies \citep{Venditti.2023}.
We note that the prism spectroscopy is taken on component A of \galname only; its component B resides $\sim$0\farcs2 (\ie $\sim$1~kpc proper) to the South East (see Sect.~\ref{subsect:morph}).
We carry out photometry for A and B separately, and find that they have comparable photo-$z$ estimates: $z^{\rm A}_{\rm phot}=8.35\pm0.29$ and $z^{\rm B}_{\rm phot}=8.93\pm0.91$, both 
in good agreement with the spectroscopic redshift measured for A (see Sect.~\ref{subsect:photoz}).
Our photometry for component B shows a blue UV continuum comparable to A's, but with a much smaller F444W/F200W flux density ratio (see Sect.~\ref{subsect:photom}).
Assuming that the different color is driven by ${\rm EW}_{\OIII}$, we estimate that B's ${\rm EW}_{\OIII}$ is a factor of $\sim$2 lower than A's, indicating that B's ISM is more chemically pristine. B's SED shape might allow for an \HeII line with higher EW, but we caution that B's photo-$z$ carries substantial uncertainties.
Simulations predict that Pop III stars can continue to form if there still are clumps of primordial gas not yet polluted by metal-enriched outflows from nearby star-forming galaxies or metal-loaded stellar winds from massive stars closeby \citep{Ciardi.2005,Venditti.2023}.
A deep NIRSpec/IFU observation with high spectral resolution covering both components can provide a more definitive answer to the origin of this prominent \HeII emission observed in \galname.

\begin{acknowledgments}
We thank the anonymous referee for very constructive comments that help improve the quality of this paper. XW thanks Jiangtao Li, Themiya Nanayakkara, James Trussler, and John Weaver for useful discussion.
XW is supported by the National Natural Science Foundation of China (grant 12373009), the CAS Project for Young Scientists in Basic Research Grant No. YSBR-062, the Fundamental Research Funds for the Central Universities, the Xiaomi Young Talents Program, and the science research grant from the China Manned Space Project.
This work is based on observations made with the NASA/ESA/CSA James Webb Space Telescope, associated with the program JWST-DD-2767.
The data presented in this article were obtained from the Mikulski Archive for Space Telescopes at the Space Telescope Science Institute, which is operated by the Association of Universities for Research in Astronomy, Inc., under NASA contract NAS 5-03127 for JWST.
The specific observations analyzed in this work can be accessed via \dataset[DOI:10.17909/2dxj-z303]{https://doi.org/10.17909/2dxj-z303}.
JG acknowledges support from the Youth Innovation Promotion Association of the Chinese Academy of Sciences (No. 2022056).
MO acknowledges support by JSPS KAKENHI grants JP20H00181, JP20H05856, and JP22H01260.
\end{acknowledgments}

\software{
\bagp \citep{Carnall.2018},
CLOUDY \citep{ferlandCLOUDY90Numerical1998},
\emc \citep{foreman-mackeyEmceeMCMCHammer2013},
\galfit \citep{2002AJ....124..266P},
the JWST Science Calibration Pipeline \citep{bushouse_2023_8067394},
MAPPINGS \citep{Sutherland.1993},
\msa \citep{https://doi.org/10.5281/zenodo.7579050},
\ppxf \citep{Cappellari.2022},
TLUSTY \citep{Hubeny.2011,Hubeny.2021}.
}

\bibliography{reference,xinlib}{}
\bibliographystyle{aasjournal}

\end{document}